%% file: main.tex
\documentclass[acmtog, nonacm]{acmart}

\AtBeginDocument{%
  \providecommand\BibTeX{{%
    \normalfont B\kern-0.5em{\scshape i\kern-0.25em b}\kern-0.8em\TeX}}}

\acmJournal{TOG}

\usepackage{booktabs}
\usepackage{savesym}
\savesymbol{zifour@default}
\savesymbol{zifour@scaled}
\usepackage{enumerate}
\usepackage{multirow} 
\usepackage[normalem]{ulem}
\usepackage{inconsolata}

\usepackage{algorithm}
\usepackage{algpseudocode}

\usepackage{listings}
\definecolor{lstbg}{gray}{0.975}
\definecolor{lstrule}{gray}{0.65}
\definecolor{lstnum}{gray}{0.55}
\definecolor{lstkw}{RGB}{0,51,153}
\definecolor{lststr}{RGB}{153,51,0}
\definecolor{lstcmt}{RGB}{96,112,96}
\definecolor{lstemph}{RGB}{122,0,122}

\lstdefinestyle{appbase}{
  basicstyle=\ttfamily\scriptsize,
  columns=flexible, keepspaces=false, tabsize=4,
  breaklines=true, breakatwhitespace=true, breakautoindent=true, breakindent=0pt,
  numbers=left, numberstyle=\tiny\color{lstnum}, numbersep=6pt, stepnumber=5, firstnumber=1,
  frame=tb, rulecolor=\color{lstrule}, framerule=0.4pt, framesep=4pt,
  backgroundcolor=\color{lstbg},
  xleftmargin=2em, framexleftmargin=2em,
  aboveskip=8pt, belowskip=4pt, captionpos=b,
  showstringspaces=false, extendedchars=true,
  literate={±}{{$\pm$}}1 {°}{{$^{\circ}$}}1 {—}{{--}}1 {–}{{-}}1 {≈}{{$\approx$}}1
}

\lstdefinestyle{apppython}{style=appbase, language=Python,
  keywordstyle=\color{lstkw}\bfseries, stringstyle=\color{lststr}, commentstyle=\color{lstcmt}\itshape,
  emph={Synthesis,Verification}, emphstyle=\color{lstemph}\bfseries,
  morekeywords={self}}
\lstdefinestyle{apptext}{style=appbase, language={}, numbers=none, xleftmargin=0.5em, framexleftmargin=0.5em}

\makeatletter
\AtBeginDocument{\let\c@lstlisting\c@figure}
\makeatother
\lstdefinestyle{appdiff}{style=apptext, numbers=none,
  moredelim=[l][\color{lststr}]{-},
  moredelim=[l][\color{lstcmt}]{+}}

\input{macros.tex}

\begin{document}

\title[\methodname: Human-LLM Co-design of Reusable Spatial Concepts for Layout Generation]{\methodname: Human-LLM Co-design of Reusable Spatial Concepts for Layout Generation}

\author{R. Kenny Jones}
\affiliation{%
  \institution{Stanford University}
  \country{USA}}
\author{Jiajun Wu}
\affiliation{%
  \institution{Stanford University}
  \country{USA}}
\author{Maneesh Agrawala}
\affiliation{%
  \institution{Stanford University}
  \country{USA}}

\begin{abstract}
\input{00-abstract.tex}
\end{abstract}

\input{figs/teaser_fig}

\maketitle

\input{01-intro}
\input{02-related}

\input{03-overview}

\input{04-concept_codesign}

\input{05-concept_graph}

\input{06-results}

\input{07-extensions}
\input{08-conclusion}

\bibliographystyle{ACM-Reference-Format}
\bibliography{main}

\pagebreak

\appendix

\input{appendix}

\end{document}

%% file: macros.tex
\usepackage{xspace}
\usepackage{cleveref}

\newcommand{\denselist}{\itemsep 0pt\parsep=0pt\partopsep 0pt\vspace{-\topsep}}

\newcommand{\methodname}{\text{Co-DSL}\xspace}

\newcommand{\nodename}{Group\xspace}
\newcommand{\nodenames}{Groups\xspace}

\newcommand{\edgename}{Relation\xspace}
\newcommand{\edgenames}{Relations\xspace}

\newcommand{\metaprog}{generative layout program\xspace}
\newcommand{\metaprogs}{generative layout programs\xspace}

%% file: 00-abstract.tex
We present \methodname, a framework in which users and LLMs co-design reusable spatial concepts for layout generation.
In the first stage, our system allows users to create a library of personalized operations.
To add a concept, a user provides a description, which the LLM converts into an initial implementation.
The user can then observe the concept's execution behavior over example scenarios and provide lightweight feedback to the LLM, which revises the concept's implementation logic.
In the second stage, given a library, an LLM can author a layout program that meets a new layout request from a user.
Executing a layout program instantiates a concept-graph, a declarative composition of library concepts.
In this way, we require the LLM only to specify which concepts should hold, while we use a solver to search for a layout that satisfies them.
Across five spatial layout domains, we find that \methodname libraries let LLMs capture layout patterns that are challenging for prior fixed-operation systems, and demonstrate that our concept-graph representation is needed for LLMs to reliably compose concepts together.
We further find that \methodname lets LLMs author \metaprogs that expose high-level parametric controls users can manipulate to explore a distribution of related layouts without re-prompting.

%% file: figs/teaser_fig.tex
\begin{teaserfigure}
    \centering
    \includegraphics[width=\textwidth]{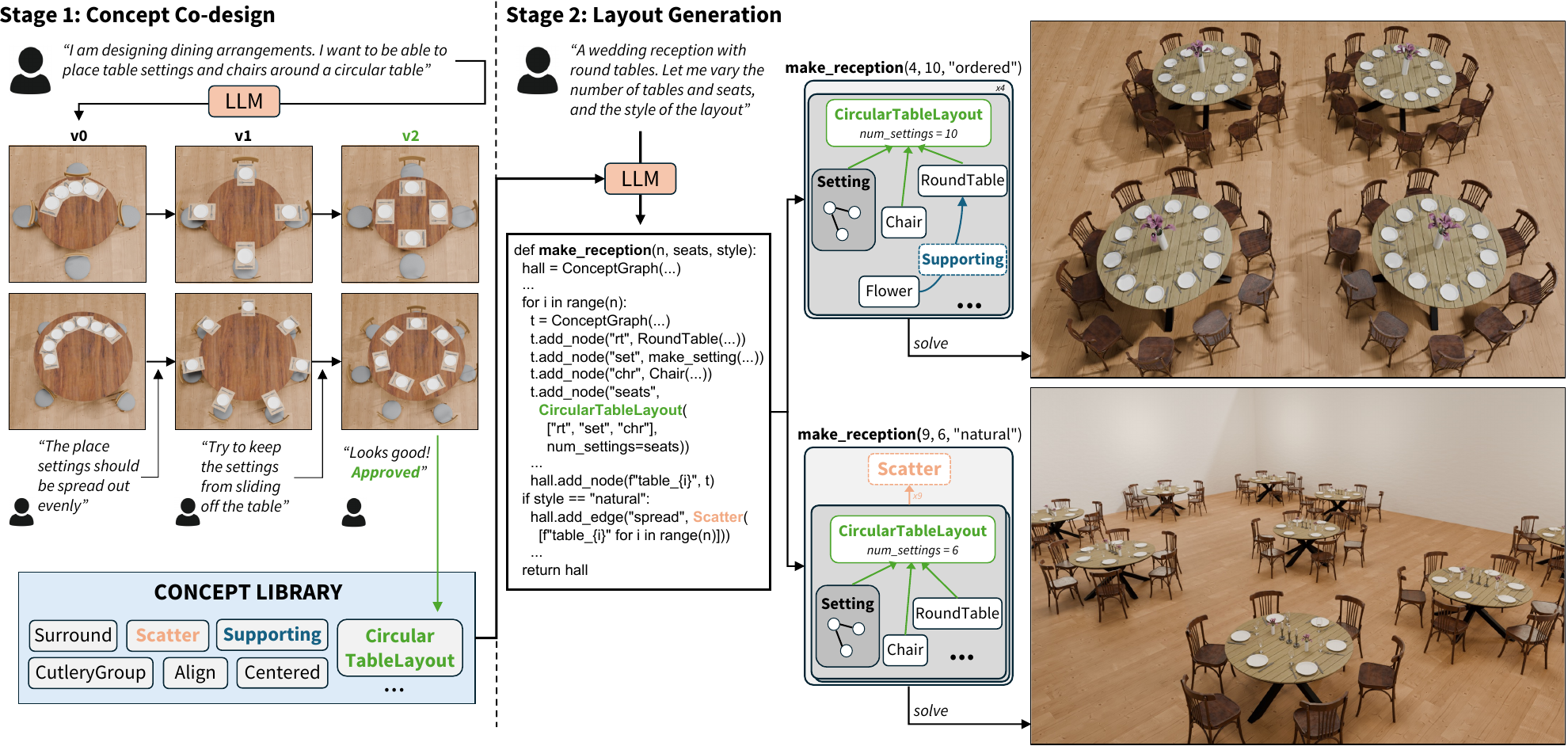}
    \caption{
We propose \methodname, a framework that helps LLMs generate spatial layouts that align with a user's intent.
In the first stage (left), a user collaborates with an LLM to co-design reusable spatial concepts, providing high-level feedback until each concept behaves as intended.
In the second stage (right), given a new layout request, the LLM authors a program that composes library concepts.
Using \methodname, we find that LLMs can author parameterized layout programs, where different parameterizations instantiate concept-graphs that generate different layouts, letting the user explore design variations without re-prompting.
 }
    \label{fig:teaser}
\end{teaserfigure}

%% file: 01-intro.tex
\section{Introduction}
\label{sec:intro}

The general capabilities of LLMs allow them to help with spatial layout tasks from furnishing a room, to setting a table, or assembling parts together to form objects.
Yet LLMs are also unreliable, often producing layouts that violate important parts of a user's specification.
A promising mitigation strategy is to frame layout generation as an LLM program synthesis task,
where LLMs write programs in a domain-specific language (DSL), and the programs can be executed or solved to produce a layout.
When the DSL contains operations that match what the prompt asks for, LLMs are typically successful in expressing user-requested spatial relationships~\cite{yang2024holodeck, Ma2025mover}.

The downside of this paradigm is that the operations of the language are fixed in advance.
Existing AI layout systems rely on expert-designed DSLs that aim to provide utility across a variety of layout domains.
As a result, their one-size-fits-all operation sets are not guaranteed to be well-suited for the particular kinds of layouts a new user might want to create. 
Users tend to think about layout specifications in terms of recurring, high-level patterns:
how to position cutlery around a dining table, how living room sofas orient towards a television, how to place multiple monitors on a desk. 
Expressing such a pattern through general-purpose, relational operations requires carefully coordinating many coupled decisions, a process that must be repeated every time the pattern is needed.
It is easy for LLMs to make mistakes in this low-level composition and produce layouts that satisfy fragments of the prompt while missing the intended pattern.

To address this gap, we propose \methodname, a framework in which a user and an LLM collaboratively design a library of programmatic spatial concepts that capture the reusable patterns the user cares about.
\methodname operates in two stages: (1) \textit{concept co-design} and (2) \textit{layout generation}.
During the concept co-design stage, the user works with the LLM to concretize domain-specific patterns, such as \texttt{CircularTableLayout}, into reusable, executable concepts (Figure~\ref{fig:teaser}, left). 
During the layout generation stage, the user provides a prompt, and the LLM composes concepts from the co-designed library to author a program that generates a layout to meet the request.
With \methodname, LLMs can author layout programs that not only better match a user's intent, but also expose interpretable parameters that support layout exploration (Figure~\ref{fig:teaser}, right).

In Stage 1, concepts are created through an iterative co-design workflow in which the user and the LLM take on complementary roles: the user reasons about how the concept should behave and the LLM reasons about how the concept should be implemented.
First, the user describes a concept in natural language, and the LLM proposes an initial implementation.
Then, the user inspects the concept's behavior over \textit{application scenarios}, small example settings in which the concept is applied.
The user provides the LLM with lightweight feedback, and the LLM revises the implementation until the concept's behavior matches the user's intent.
In this way, \methodname allows a user to encode their design preferences as reusable, interpretable abstractions, without authoring any DSL code by hand.

In Stage 2, a user prompts for a new layout, and the LLM must compose library concepts to produce it.
LLMs face two main challenges in this process. 
First, concept applications must be coordinated so that satisfying one concept does not violate relations established by another (e.g., rotating a sofa to face a TV may break its alignment with a wall).
Second, when a set of concepts cannot be jointly satisfied, the LLM must identify which of them is responsible for the conflict.
We therefore have the LLM author a layout program that instantiates a \textit{concept-graph}: a declarative specification of which library concepts should hold over which scene entities.
We introduce a solver that schedules concept applications and revisits unsatisfied concepts to search for a layout that satisfies the entire graph.
This allows the LLM to specify which relationships should hold without having to direct the search itself.
When the solver cannot find a satisfying layout, it reports unsatisfied concepts as interpretable, concept-level errors that the LLM uses to revise the layout program.

We apply \methodname across five spatial layout domains spanning three application areas: 3D room layouts, 2D layered arrangements, and 3D shape assemblies.
We show that co-designed concepts can encode domain-relevant reusable logic that captures spatial patterns that are difficult to express reliably using only general-purpose relational operations.
To evaluate the downstream benefit offered by \methodname, we develop libraries co-designed by an expert, and compare against prior LLM-based layout systems with fixed operation sets.
We find that using \methodname concepts allows LLMs to produce complex layouts that better satisfy requested spatial relationships, while maintaining physical validity, and are more often preferred by human raters.
Ablations show that providing an LLM with co-designed concepts is insufficient for reliable layout generation: our concept-graph representation, solving procedure,
and concept-level program repair substantially improve its ability to effectively compose these concepts into satisfying layouts.
Additionally, we conduct a non-expert user study, which shows that people can efficiently use \methodname to co-design concepts that match their intent without writing code.
Finally, we demonstrate new applications enabled by our framing:
(i) \metaprogs whose high-level controls allow users to explore design variations without re-prompting and
(ii) extensible concept libraries that users can grow on demand when generated layouts reveal missing concepts.
We will release code upon publication.

\noindent
In summary, our contributions are:
\begin{enumerate}[(i)]
\denselist
\item \methodname, a framework that casts LLM layout generation as program synthesis over user-specific libraries of reusable, executable spatial concepts.  
\item For concept co-design, an interactive workflow in which users guide an LLM to implement concepts that behave as they intend, without manually authoring code.
\item For layout generation, a declarative concept-graph representation, together with a solver and program repair loop, that enables LLMs to reliably compose concepts.
\end{enumerate}

%% file: 02-related.tex
\section{Related Work}
\label{sec:rel_work}

\input{figs/overview_fig}

\textbf{Procedural and constraint-based layout generation.}
Procedural layout generation has long demonstrated that explicit symbolic structure enables efficient, controllable synthesis.
Some work encodes domain knowledge through grammars or production rules, for example to generate buildings and cities~\cite{wonka2003instant,muller2006procedural,MetropolisProceduralModeling}.
Other methods instead encode design guidelines as costs or constraints, which are used in search or optimization to produce new arrangements~\cite{Xu2002ConstraintPlacement,yu2011MakeItHome,InteractiveFurnitureLayout,LARJ,Weiss2019PositionBased,infinigen2024indoors}.
Many of these systems rely on structured representations, for instance by organizing layout elements into hierarchies and symmetry groups, or by representing relationships between elements as graphs~\cite{sigg_structureAware:13,wang2011,fisher2011}.
Following this lineage, we introduce a concept-graph intermediate representation that allows an LLM to flexibly specify how executable spatial concepts relate to layout elements.
Rather than relying on rules defined in advance by experts, \methodname allows a user and an LLM to co-design reusable operations without requiring the user to author any DSL code by hand.

\textbf{Data-driven layout generation.}
Another line of work learns to model layout distributions from data.
Early methods fit structured probabilistic models or grammars to small-scale scene datasets~\cite{fisher2012,liu_scenegrammar}.
This was followed by deep learning approaches that train on larger datasets using convolutional, autoregressive, or diffusion-based architectures~\cite{DeepSynthSIGGRAPH2018,atiss,diffuscene,instructscene}.
Related approaches investigate how neural networks could be adapted to model more structured visual representations, such as hierarchical graphs or even programs~\cite{GRASS,GRAINS,StructureNet,jones2020shapeAssembly,wang2019planit,Para2021ConstraintGraphs}.
Others look towards pretrained visual models for another prior that can guide the creation of spatial arrangements~\cite{GraphDreamer,GALA3D}.
Some methods further augment data-driven generators with differentiable expert-knowledge constraints, such as ergonomic guidelines~\cite{layoutenhancer}.
We view such regularizers as complementary to \methodname's user co-designed concepts.
Through learning, these methods are able to capture general-purpose spatial patterns commonly found in a data distribution.
In contrast, \methodname uses no task-specific training data, instead users guide an LLM in encoding the spatial patterns they care about as reusable code.

\textbf{LLM-based layout generation.}
Early text-based scene generation methods relied on expert rules~\cite{WordsEye} or learned priors over a fixed set of relations~\cite{Chang2014SpatialKnowledge}.
Following the successes of LLMs, a large body of recent work has further explored how to employ them in translating from text prompts to layouts.
In the simplest case, methods ask the LLM to directly emit parameterized entities to form a layout, for instance as a set of boxes~\cite{feng2023layoutgpt,SceneTeller}.
More expressive variants ask LLMs to author procedural code~\cite{zhang2024scene}, which can even admit automatic error correction~\cite{gumin2025proceduralsceneprogramsopenuniverse}.
Another family of works pairs LLMs with expert-designed, general-purpose constraints: the LLM declares atomic entities and their low-level relationships, and a solver searches for a satisfying layout~\cite{yang2024holodeck,aguina2024open,littlefair2025flairgpt,sun2025layoutvlmdifferentiableoptimization3d,idesign}.
Beyond scene layout, related approaches have used LLMs to author shape and CAD programs~\cite{BlenderLLM,CADMLLM,Text2CAD}.
Very recently, `agentic' systems have been proposed that equip LLMs with tools, verification procedures, and orchestrating harnesses which allow them to scale inference-time compute through repeated layout refinements~\cite{SAGE,SceneWeaver}.
Across these approaches, the representations and operations available to the LLM are fixed in advance and designed for general use.
We find that LLMs using such operation sets often fail to capture the spatial patterns a user cares about, while \methodname allows these patterns to be reliably encoded as reusable operations.

\textbf{Library learning and tool creation.}
Library learning methods aim to improve program synthesis performance by discovering reusable abstractions that capture structure shared across programs or tasks~\cite{ellis2021dreamcoder,LAPS,babble,jones2021shapeMOD,jones2023shapecoder,grand2024lilo}.
These abstractions allow future problems to be solved using higher-level operations rather than the original low-level primitives.
Relatedly, LLMs can also create tools using their prior knowledge and through feedback sourced from interacting with an environment~\cite{voyager,CREATOR,hu2024scenecraft,TroVE,ReGAL}.
Closer to our setting, several systems have explored discovering abstractions from a user-supplied set of visual examples or function descriptions~\cite{shapelib, tam2024scenemotifcoder}.
However, whereas these works remove the user from the loop after the initial setup, \methodname supports iterative, interactive co-design where the user guides the LLM toward a concept implementation with agreed-upon behavior that the LLM can flexibly reuse.

\textbf{Interactive design systems.}
A large body of work has studied interactive graphics systems that allow users to steer design processes by inspecting and responding to visual outputs.
Some methods help users complete or revise a particular design through rounds of suggestions and feedback~\cite{yu2016,InteractiveFurnitureLayout,Chat2Layout}.
Other systems help with exploration by presenting users with alternatives from a parameterized design space~\cite{DesignGalleries,Talton2009CollaborativeDesignSpaces}.
Related approaches have instead allowed user interaction to guide the development of an underlying procedure, for example by specifying program behavior through input-output examples or by directly manipulating visual outputs~\cite{FlashFill,SketchNSketch,Lipp2008InteractiveGrammars}.
\methodname similarly relies on user interaction to translate design intent into executable behavior, but does so at the level of reusable spatial concepts.
In our paradigm, the user specifies and judges the behavior of each concept, while the LLM implements it as executable code.
Together, the resulting concepts form a domain-specific library that the LLM can use to meet new design requirements.

%% file: figs/overview_fig.tex
\begin{figure*}[t!]
\centering
\includegraphics[width=\textwidth]{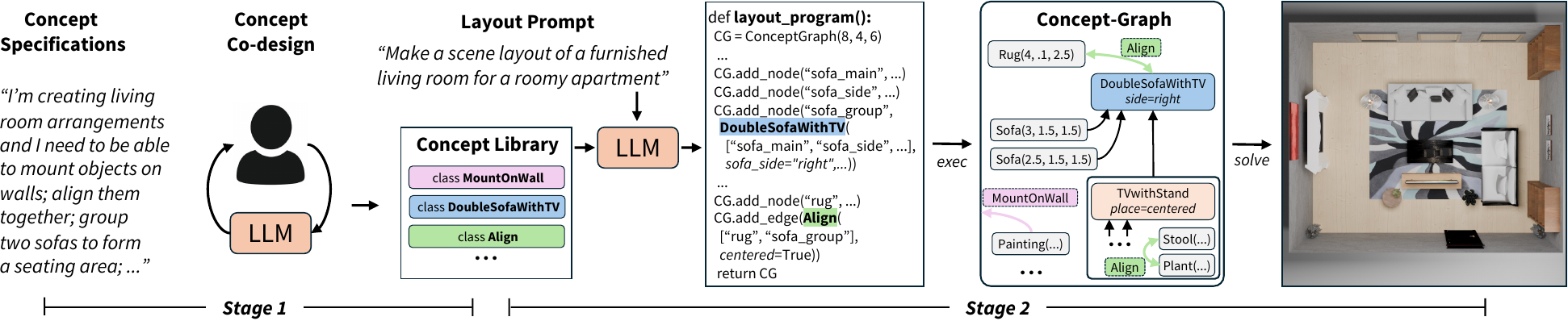}
\caption{Overview of \methodname.
During concept co-design (left), a user comes up with concept specifications and then collaborates with an LLM to co-design a library of spatial concepts.
During layout generation (right), we prompt an LLM to author a layout program from a text description.
Layout programs compose library concepts together to form concept-graphs, which we can solve to find a layout that satisfies the constraints induced by the high-level concepts.} 
  \label{fig:overview}
\end{figure*}

%% file: 03-overview.tex
\input{figs/exec_fig}

\pagebreak

\section{\methodname Overview}
\label{sec:overview}

LLM layout generation involves two coupled challenges: the LLM must interpret a user's intended spatial patterns and then produce a layout that jointly satisfies them.
As summarized in Figure~\ref{fig:overview}, \methodname addresses these challenges with separate \textit{concept co-design} and \textit{layout generation} stages.

During Stage 1, concept co-design (Section~\ref{sec:codesign_algo}), the user identifies recurring concepts that would be useful in a target modeling domain, such as mounting an object on a wall, arranging two perpendicular sofas around a TV, or aligning two objects.
The user describes each concept in natural language, and the LLM implements it as an executable class, such as \texttt{MountOnWall}, \texttt{DoubleSofaWithTV}, or \texttt{Align}.
Our co-design interface allows the user to inspect the concept's behavior in \textit{application scenarios}, LLM-authored examples that apply the concept to a few entities under varying initial conditions.
The user provides feedback, which the LLM uses to revise the implementation, until the behavior matches their intended semantics.

During Stage 2, layout generation (Section~\ref{sec:compose_concepts}), the user provides a new layout prompt, such as \textit{``a furnished living room for a roomy apartment.''}
The LLM uses the library to author a layout program that instantiates a concept-graph, a declarative specification of which concepts should hold over which entities.
Our solver then uses the execution logic of those concepts to search for a layout that satisfies the graph.

\paragraph{Layout domain requirements. }
\label{sec:domain_reqs}
\methodname targets spatial domains in which entities with attributes must be arranged within a bounded coordinate space while satisfying geometric and semantic relationships.
To operate in such settings, we require a small amount of \textit{base domain knowledge} that is shared across concepts in a domain.
This includes
\textbf{spatial bounds and frames}, such as representing a living room using a 3D box in a Y-up coordinate system; 
an \textbf{entity model}, defining atomic entity types, such as couches and TVs, with attributes such as size, position, and facing direction; 
\textbf{scene initialization}, specifying how initial layouts are sampled, such as randomly placing objects on the ground within bounds;
and \textbf{base validity checks} that all valid layouts should obey, such as requiring objects to remain in bounds, not be floating, and avoid collisions.
We provide further details in Appendix~\ref{sec:app_base_domain}.

%% file: figs/exec_fig.tex
\begin{figure*}[t!]
\centering
\includegraphics[width=\textwidth]{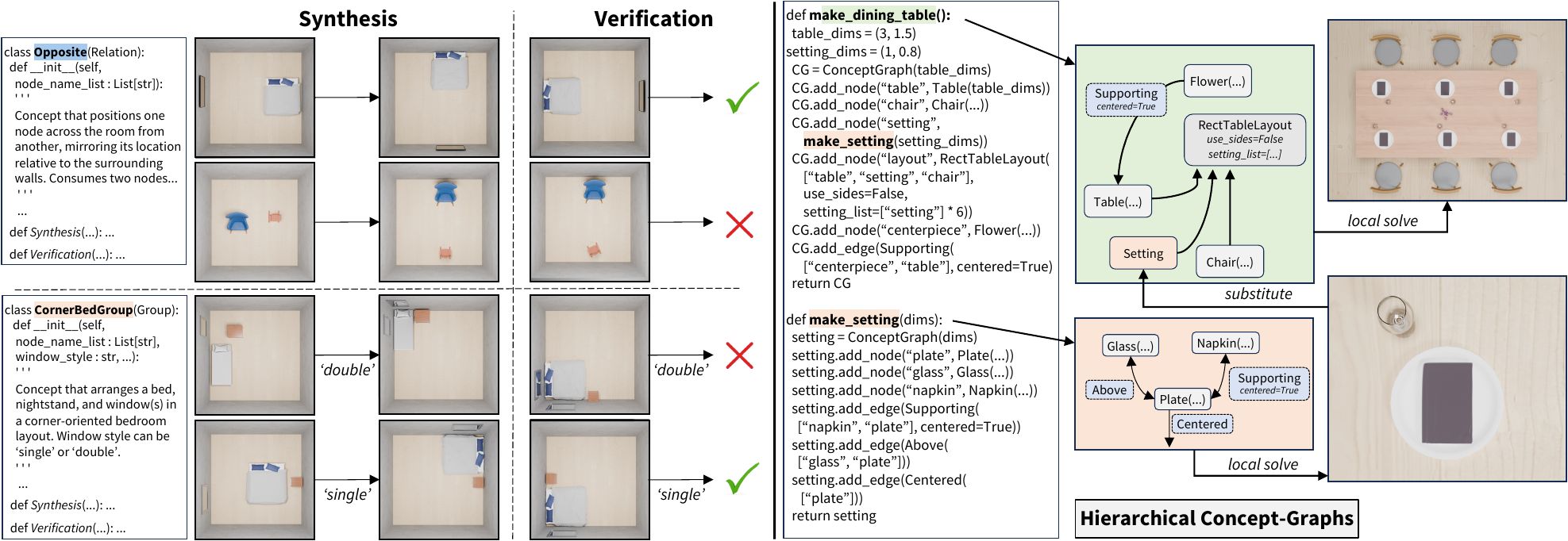}
\caption{
\methodname concepts and concept-graphs.
(\textit{Left}): each concept is implemented as a parameterized \nodename or \edgename class with dual execution modes: 
\textit{synthesis} logic transforms an input layout into an output layout, and \textit{verification} logic checks whether a static layout satisfies the concept's semantics.
(\textit{Right}): executing a layout program instantiates a concept-graph, a declarative specification of which concepts should hold over which entities. 
A solver searches for a layout satisfying this specification, treating layouts from child graphs as single units within their parent.
} 
  \label{fig:exec_engine}
\end{figure*}

%% file: 04-concept_codesign.tex
\input{figs/codesign_fig}

\section{Stage 1: Human-LLM Concept Co-design}
\label{sec:codesign_algo}

\textbf{Concept representation. }
\label{sec:concepts}
\methodname represents each concept as an executable class with parameters that control its behavior (Figure~\ref{fig:exec_engine}, left).
Some concepts specify relationships between entities; we represent these as \edgename classes.
For example, \texttt{Opposite} places two objects so that they face one another while remaining separated.
Other concepts organize entities into logical units; we represent these as \nodename classes.
For example, \texttt{CornerBedGroup} arranges a bed with related objects, and its \texttt{window\_style} parameter controls how many windows are created and how they are placed.

Each concept supports two execution modes: \textit{synthesis} and \textit{verification}.
The \textit{Synthesis()} function takes a layout as input and modifies the entities referenced by the concept, producing an output layout that better matches the concept's intended spatial pattern.
The \textit{Verification()} function takes a layout as input and returns a boolean indicating whether the concept is satisfied.
As a concept typically admits many valid layouts, a synthesis function can produce different transformations depending on the input layout state or internal stochastic choices.
This allows repeated applications to explore alternative layouts that pass the corresponding verification check.

\textbf{Concept initialization. }
The user begins with a \textit{concept specification}: a name, whether the concept is a \nodename or \edgename, a natural-language description of its intended behavior, and any parameters that control that behavior.
From this specification, the LLM proposes a concept interface consisting of a class template, constructor, and docstring that describes the expected behavior.
The user can accept this interface or provide feedback to revise it.
Once the interface is accepted, the LLM proposes initial implementations of the synthesis and verification functions.

\textbf{Application scenarios. }
To judge a concept's implementation without reading code, the user needs to observe its behavior in settings where the concept could be used.
\methodname provides these settings through \textit{application scenarios}: small programs that apply a concept to a few entities under varying initial conditions.
Each scenario has a short description of a target arrangement, which the user can provide or ask the LLM to propose.
Given this description and the base domain knowledge (including available entities and scene initialization logic), the LLM authors a program that creates the required entities, places them in the scene, and specifies how the concept applies to them.
For instance, the \texttt{chairs\_evenly\_placed()} scenario in Figure~\ref{fig:codesign} applies \texttt{Surround} to a table and three chairs.
We direct the LLM to introduce random variation into scenario programs so that repeated executions expose concept behavior across different initial layouts and parameterizations.
The user can inspect how synthesis transforms input layouts into output layouts and how verification judges both valid and invalid arrangements.

\textbf{Co-design interface. }
We develop an interactive co-design interface whose views each expose a different facet of a concept's current implementation.
The \textbf{interface view} contains the class constructor and docstring.
The \textbf{scenario view} presents application scenarios.
The \textbf{synthesis view} and \textbf{verification view} visualize applications of the concept's synthesis and verification functions.
Once satisfied, the user adds a concept to the library from a \textbf{confirmation view}.

\textbf{User guidance. }
A user can provide feedback from any view, which directs the LLM to revise the corresponding part of the concept.
Text-based feedback continues the existing conversation with the LLM, directing it to revise its previous implementation.
Users can also point the LLM towards an existing library concept, which is added to the prompt as an in-context example.
The verification view additionally supports an \textit{auto-fix} action.
From this view, a user can inspect layouts that the current verifier judges as \textit{valid} or \textit{invalid}.
If the user disagrees with these votes, they can record their own label, forming a small validation set.
The \textit{auto-fix} procedure gives the LLM the current verification function and the layouts on which the user disagreed, and directs the LLM to revise the verifier logic to match the user's judgments.
This allows a user to guide a concept's development even when describing the exact behavior they want is challenging.

\subsection{Walkthrough of Concept Development}
\label{sec:CD_walkthrough}

We provide a walkthrough of how iterative concept co-design may unfold for a representative concept, \texttt{Surround}, in the dining domain (Figure~\ref{fig:codesign}).

The user begins in the \textbf{interface view} by providing a specification:
\texttt{Surround} is a \edgename that constrains a central entity to be in the middle of a group of outer entities, modulated by \texttt{distance} and \texttt{spacing\_mode} parameters.
The LLM uses this input to propose an interface.
The user checks the class interface, and then directs the LLM to implement the initial synthesis and verification functions.

Navigating to the \textbf{scenario view}, the user directs the LLM to author two application scenarios: one where ``three chairs are evenly placed around a table'' and one where ``four tables are unevenly surrounding a central fountain.''
For each description, the LLM writes a scenario program, and the scenario view shows the initial layouts to which the concept will be applied.

The user might then move on to the \textbf{synthesis view} to inspect how the synthesis behavior transforms the initial scenes from each application scenario.
The user observes that although generated layouts place the chairs near the table, the chair orientations are incorrect.
The user then provides guidance to the LLM by describing the intended orientation behavior, and the LLM updates the synthesis implementation accordingly.

Changing to the \textbf{verification view}, the user looks at how the concept's verification function judges layouts as either valid or invalid.
The user might notice that the verification logic is producing false positives because it does not adequately check a desired condition: there are \textit{valid} votes for some arrangements where the fountain lies outside of the convex hull formed by the surrounding tables.
The user relabels these layouts as \textit{invalid} and invokes the \textit{auto-fix} action, guiding the LLM on how to update the verification logic.

At this point, the user might return to the \textbf{synthesis view} and notice that for the application scenario with unequal spacing, the synthesis function still places the outer entities at nearly uniform radial distances from the central entity.
They can request another change to the synthesis function so that this parameter setting produces their intended diversity in placement behavior.

Finally, the user can navigate to the \textbf{confirmation view}, which summarizes whether an in-development concept is ready to be added to the library.
It exposes diagnostics such as coverage across application scenarios, verifier agreement on user-labeled validation set examples, and how often synthesized outputs also satisfy verification.
Once satisfied, the user clicks confirm to approve the concept's addition to the library.

%% file: figs/codesign_fig.tex
\begin{figure*}[t!]
\centering
\includegraphics[width=\textwidth]{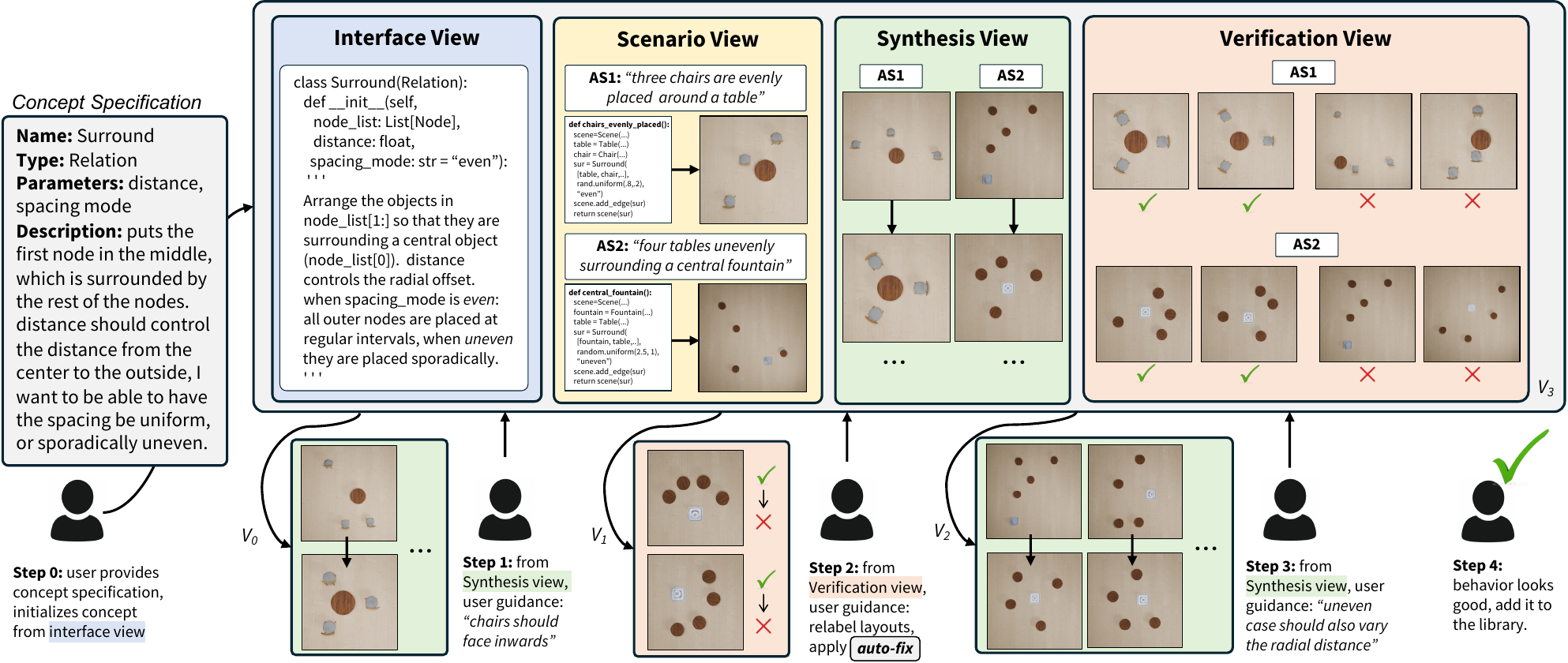}
\caption{
Co-designing a concept. 
The user starts with a \textit{concept specification} (left), which the LLM uses to initialize the concept.
The user then navigates diagnostic views to inspect the concept's behavior.
The interface view shows the class definition, the scenario view shows application scenarios, the synthesis view shows \textit{Synthesis()} transformations, and the verification view shows \textit{Verification()} predictions.
The user can provide feedback from any view to guide the LLM in making revisions.
We show an example guidance sequence a user might provide while developing a \texttt{Surround} \edgename (see Section~\ref{sec:CD_walkthrough}).
}
\label{fig:codesign}
\end{figure*}

%% file: 05-concept_graph.tex
\section{Stage 2: Layout Generation with Concept-Graphs}
\label{sec:compose_concepts}

The goal of \methodname's layout generation stage is to produce a layout that meets the requirements of a user's \textit{layout prompt} (Figure~\ref{fig:overview}, right).
Given a new request, \methodname instructs an LLM to author a \textit{layout program} using concepts from a co-designed library.
Executing a layout program instantiates a \textit{concept-graph}, a declarative specification of which concepts should hold over which entities.
We can then use a solver to search for a layout in which every concept in the graph is satisfied.
This division allows the LLM to focus on deciding which concepts should hold, while the solver figures out how to meet this specification.

Programs can make randomized or parameterized choices, so different executions of the same program may instantiate different graphs, for example varying the number of place settings or the arrangement pattern of a table.
This gives the solver alternatives to explore when one instantiated graph does not yield a satisfying layout.
When the solver cannot find any satisfying arrangement from a layout program, the concepts it failed to satisfy are reported back to the LLM, which revises the program logic.
Our procedure therefore has three nested loops: an outer loop that revises layout programs through LLM repair, a middle loop that searches over concept-graphs instantiated by one program, and an inner loop that searches for a layout satisfying the graph.

\subsection{Authoring Layout Programs}
\label{sec:layout_authoring}

To author a layout program, \methodname provides the concept library to the LLM as an API.
For each concept, the LLM receives its class interface, consisting of the constructor signature and docstring produced during co-design.
The prompt also describes the base domain knowledge, explains how concept-graphs are constructed, and optionally includes user-approved layout programs as in-context examples.
The LLM does not see the synthesis or verification logic of any concept, so it composes concepts based on their documented behavior and leaves the execution of that behavior to the solver.
Appendix~\ref{sec:app_walkthrough} walks through the authoring prompt and an example run.

\subsection{Concept-Graphs}
\label{sec:concept_graphs}

A concept-graph is a structured specification of the entities in a layout and the spatial constraints they must satisfy.
It acts as an intermediate representation, allowing LLMs to declaratively compose \nodename and \edgename library concepts.
In this graph, nodes represent base-domain entities, \nodename concepts, or child concept-graphs, while \edgename concepts form edges between those nodes.
The graph also stores parameters for the bounded coordinate space in which the layout should be generated.

Figure~\ref{fig:exec_engine} (right) shows a layout program instantiating a concept-graph.
The \texttt{Above} concept introduces an edge relationship over two entity nodes, a \texttt{Plate} and a \texttt{Glass}.
A \nodename node can represent a logical collection of layout elements, which other concepts can reference as a single unit.
Similarly, a child concept-graph is represented as a single node within its parent.
For instance, the concept-graph produced by \texttt{make\_setting()} becomes a single node in the parent graph instantiated by \texttt{make\_dining\_table()}.
The \texttt{RectTableLayout} \nodename concept takes this setting node as input, creates $N$ copies of its entities, and arranges the copied settings around a table.
The concept-graph thus specifies which concepts should hold and how they relate, without requiring the layout program to schedule \textit{Synthesis()} and \textit{Verification()} calls directly.

\subsection{Solving Concept-Graphs}
\label{sec:graph_solving}

\input{algs/solve_cg}

Algorithm~\ref{algo:solve_cg} shows how we search for a layout given a layout program $P$ and a time budget $\tau$.
It has two procedures: \textproc{Solve} searches over executions of $P$, while \textproc{SolveLocal} searches for a layout satisfying an instantiated concept-graph.
We measure progress with a graph error that aggregates base validity failures, unsatisfied concept verifications, and errors propagated from child concept-graphs.
A graph is solved when its error is zero.

\textproc{Solve} repeatedly executes $P$ to instantiate a concept-graph.
For each graph, \textproc{Solve} calls \textproc{SolveLocal} to obtain a candidate scene and its error, retaining the best scene found across attempts.
The solver returns the first zero-error scene it finds, or the lowest-error scene discovered within $\tau$.

Within each attempt, \textproc{SolveLocal} uses the graph's hierarchy to solve child graphs before combining their layouts in the parent.
We first initialize an empty scene using the graph's coordinate-space parameters.
We recursively solve each child, take its output layout, and treat the resulting sub-scene as a single node in the parent.
The parent can transform each sub-scene as a unit, but its internal arrangement remains fixed.
We maintain a memoized cache of the best scene found for each local concept-graph, allowing solved subgraphs to be reused across repeated solve attempts.

\subsubsection{Interleaving Synthesis and Verification}
\label{sec:local_solve}

Once the child layouts have been inserted,
\textproc{SolveLocal} runs a synthesis pass over the concepts of the graph.
We iterate over concepts using a randomized dependency-aware traversal, selecting only concepts whose referenced entities already exist in the scene.
We apply each concept's \textit{Verification()} function, and if it returns \textit{False}, we apply its \textit{Synthesis()} function to update the scene.
As described in Section~\ref{sec:concepts}, synthesis proposes a change to better satisfy the concept.
Gating such modifications on verification avoids unnecessary transformations that could disturb other relationships in the scene.
After the pass, we recompute the graph error.

If the error is nonzero, we enter a verification-driven repair loop.
Each iteration revisits the concepts that still fail verification and reapplies their synthesis functions.
After each repair pass, we recompute the error and retain the updated layout only if it improves on the best layout found locally.
The loop stops when the graph is satisfied or a pass fails to reduce the error, returning the best local layout and updating the cache.

An alternative approach would run each concept's synthesis once and then verify the resulting layout, but it cannot respond when later transformations violate relationships established by
  earlier concepts.
By interleaving synthesis and verification, our solver is able to repair these broken relationships.
Our ablations in Section~\ref{sec:res_ablation} show that removing the repair loop (\textit{w/o solver repair}) or concept-specific verification (\textit{w/o verification}) substantially reduces how often graphs are solved.

\subsection{LLM-Driven Program Repair}
\label{sec:program_repair}

In some cases, an LLM may author a layout program for which our solver cannot find a satisfying layout.
We address this with an outer correction loop that feeds concept-level failures back to the LLM.
Concretely, we provide the LLM with the current program, the original layout prompt, and the set of unsatisfied concepts or base validity errors observed during solving.
We then instruct the LLM to revise the program so that it produces satisfiable concept-graphs while remaining faithful to the prompt.
After each revision, we rerun the solver on the new program, directing the LLM to revise it again if we are unable to find a satisfying layout.
Because \methodname represents both programs and errors at the concept level, LLM-based correction remains aligned with the concepts that the user cares about, rather than requiring low-level entity manipulation or constraint repair.

%% file: algs/solve_cg.tex
\algtext*{EndProcedure}
\algtext*{EndWhile}
\algtext*{EndFor}
\algtext*{EndIf}

\begin{algorithm}[t!]
\begin{flushleft}
\begin{algorithmic}[1]
\State \textbf{Input:} program $P$, time budget $\tau$
\State \textbf{Output:} best scene $S^\star$
\Procedure{Solve}{$P, \tau$}
    \State Init $S^\star$, $e^\star \gets \infty$, cache $\mathcal{C}$
    \While{time remains before $\tau$}
        \State $G \gets \textproc{exec}(P)$
        \State $(S, e) \gets \textproc{SolveLocal}(G, \mathcal{C})$
        \If{$e = 0$} \textbf{return} $S$ \EndIf
        \If{$e < e^\star$} $S^\star \gets S; e^\star \gets e$ \EndIf
    \EndWhile
    \State \Return $S^\star$
\EndProcedure
\Statex
\Procedure{SolveLocal}{$G, \mathcal{C}$}
    \If{$\mathcal{C}.\textproc{HasPassing}(G)$} \Return $\mathcal{C}.\textproc{GetPassing}(G)$ \EndIf
    \State $S \gets G.\textproc{InitScene}()$
    \ForAll{child concept-graphs $H$ in $G$}
        \State $(S_H, \_) \gets \textproc{SolveLocal}(H, \mathcal{C})$
        \State $G.\textproc{Replace}(H, S_H)$ \hfill \Comment{Insert frozen sub-scene}
    \EndFor
    \ForAll{concepts $c$ in $G$} \hfill \Comment{Synthesis pass}
        \If{$\neg c.\textproc{Verification}(S)$}
            $c.\textproc{Synthesis}(S)$
        \EndIf
    \EndFor
    \State $e \gets G.\textproc{ComputeError}(S)$
    \While{$e > 0$} \hfill \Comment{Verification-driven repair}
        \ForAll{concepts $c$ in $G$}
            \If{$\neg c.\textproc{Verification}(S)$}
                $c.\textproc{Synthesis}(S)$
            \EndIf
        \EndFor
        \State $e' \gets G.\textproc{ComputeError}(S)$
        \If{$e' \ge e$} \textbf{break} \EndIf
        \State $e \gets e'$
    \EndWhile
    \State $\mathcal{C}.\textproc{Update}(G, S, e)$
    \State \Return $(S, e)$
\EndProcedure
\end{algorithmic}
\end{flushleft}
\caption{Solving Concept-Graphs}
\label{algo:solve_cg}
\end{algorithm}

%% file: 06-results.tex
\input{figs/domain_res_fig}

\section{Results}
\label{sec:results}

We evaluate \methodname across five spatial layout domains with concept libraries co-designed by an expert user (Section~\ref{sec:res_domains}).
We first examine whether these co-designed concepts capture spatial patterns that are difficult for LLMs to express using fixed operation sets (Section~\ref{sec:res_encoding}).
Next, we evaluate whether access to these concepts improves LLM generation of layouts from text, comparing against prior LLM program-synthesis systems (Section~\ref{sec:res_text_to_layout}).
To understand which parts of our method enable LLMs to reliably compose concepts, we ablate the concept-graph representation, program-repair procedure, and solver machinery (Section~\ref{sec:res_ablation}).
Finally, we study whether non-expert users can use \methodname to create concepts that match their intent (Section~\ref{sec:res_user_study}).

\subsection{Evaluation Domains and Concept Libraries}
\label{sec:res_domains}

We organize our five evaluation domains into three application areas.
\textbf{3D rooms} model scenes using 3D object bounding boxes with position, size, and facing direction (\textit{living rooms, bedrooms}).
\textbf{2D layered arrangements} model 3D scenes from a domain-specific 2D perspective before lifting them into 3D (\textit{dining arrangements, bookshelves}).
\textbf{3D shape assemblies} model objects as arrangements of 3D primitive parts (\textit{chairs}).
Each domain provides the base domain knowledge described in Section~\ref{sec:domain_reqs}.

For each domain, an expert used \methodname to co-design a library of concepts.
We showcase representative concepts from these domains in Figure~\ref{fig:domain_res}.
These concepts encode reusable logic that can be controlled through interpretable parameters.
For instance,
\texttt{CubbyGrid} creates arrangements of regularly placed cubby sub-spaces that can hold other objects,
while \texttt{WallGallery} can vary both the number of columns and whether wall items are regularly tiled or artistically staggered.

In total, the expert used \methodname to co-design over 90 concepts across these five domains.
On average, converting an input concept specification into a finalized library concept took an expert less than 10 minutes and less than \$1 in LLM API calls.
Free-form text guidance was provided on average three times per concept, mostly in the synthesis and scenario views.
Guidance to the verification logic mostly came through the \textit{auto-fix} procedure ($\sim$1 call per concept).
This procedure was very successful in finding improved verification functions,
each call improved verifier accuracy on the validation labels by an average of 35 percentage points and produced perfect agreement with the user in half of those cases.

\subsection{Encoding Spatial Patterns as Reusable Concepts}
\label{sec:res_encoding}

\input{figs/res_fig_concept_encoding}

We next evaluate whether reusable \methodname concepts can encode user-specified, domain-relevant spatial patterns that are difficult to express with fixed operation sets.
We compare against two recent LLM layout-generation systems designed for indoor rooms with static, expert-designed DSLs.
Holodeck~\cite{yang2024holodeck} uses programs that declare low-level spatial constraints between scene entities.
PSDL~\cite{gumin2025proceduralsceneprogramsopenuniverse} instructs an LLM to author procedural layout programs and uses error correction to repair them.

Figure~\ref{fig:concept_encoding} compares how well these methods are able to match the intent of individual user-specified layout patterns.
Each row shows a representative concept from one of our five evaluation domains.
The left column shows layouts that the user approved as matching the concept's semantics during co-design.
The following columns show how well each method can apply the requested pattern in a new layout context.
As \methodname has access to user co-designed concepts from the library, we analogously provide PSDL and Holodeck with information about the concept by appending a user-provided concept description to each prompt.

This concept-level evaluation studies how well each method is able to maintain a concept's semantics in a new layout setting.
While the fixed-operation methods are able to satisfy some parts of each request, they are often not well-aligned with the user's conceptual intent.
PSDL places nightstands out of reach of the head of the bed, orients sofas away from the TV they should face, and arranges cutlery in a jumbled, inconsistent manner. 
Holodeck often drops part of the pattern, such as omitting place settings entirely, or fails to bring related objects close enough together to be grouped visually, such as placing a side and main sofa at opposite ends of a room.

In contrast, \methodname is able to reliably reuse the library concepts and apply them in new contexts to meet the user's specifications.
We also include a \textit{w/o user guidance} condition, in which the LLM implements the concept from the user's initial specification without receiving subsequent feedback.
These examples illustrate how initial LLM implementations often encode behavior that differs from the user's expectation.
Note that as PSDL and Holodeck do not support bookshelf arrangements or 3D shape assemblies, we mark their outputs for these domains as unsupported.

\input{figs/res_fig_validity_adherence}

\input{figs/res_fig_qual_comp}

To supplement these qualitative trends, we conduct an expert analysis of whether each generated layout maintains the user-defined semantics.
We select three concepts from three domains and develop two layout prompts for each concept, yielding 18 generations per method.
When using the co-designed concepts, the expert judged that \methodname respected the pattern semantics in every generated layout.
In contrast, the expert found that the intended semantics were preserved in at most one third of the layouts generated from PSDL, Holodeck, and the \textit{w/o user guidance} ablation.
Appendix~\ref{sec:app_concept_adherence} breaks this analysis down by concept.

\subsection{Generating Layout Programs from Text}
\label{sec:res_text_to_layout}

Once a concept library has been co-designed, generating new layouts requires only a text prompt from the user (Section~\ref{sec:compose_concepts}).
Having evaluated concept-level adherence, we now test whether LLMs can use concept libraries to generate layouts that satisfy prompts with multiple interacting specifications.

\textbf{Experimental setup. }
We use 20 prompts per domain, generated by an LLM from a few in-context examples and written to exercise structural variation within each domain.
As Holodeck and PSDL cannot generate bookshelf arrangements or chairs, we study the living-room, bedroom, and dining domains.
We generate two layouts per prompt, yielding 120 generations per method across the three domains.
For fairness, we run all methods using the same base LLM (gpt-5.5), see Appendix~\ref{sec:app_comparisons} for full comparison details.

\textbf{Layout generation metrics. }
We evaluate generations along two complementary axes.
\textit{Physical validity} reports the percentage of layouts that contain no base validity errors: objects are in bounds, not floating, and do not intersect other objects.
\textit{Prompt adherence} reports the percentage of explicitly requested spatial constraints in the prompt that an LLM judges as satisfied in the generated layout.
Like \methodname, both PSDL and Holodeck explicitly seek to avoid base validity errors: Holodeck incorporates physical-validity constraints into its optimization, while PSDL uses error correction to repair the same types of violations.
This comparison therefore measures how well each method is able to jointly satisfy both prompt semantics and physical constraints.

\textbf{Comparison against prior methods. }
We plot each method's performance, split by domain, in Figure~\ref{fig:validity_adherence}.
\methodname combines high physical validity (95\% to 100\% across the three domains) with high prompt adherence (on average 90\%).
With a declarative, low-level operation set, Holodeck produces fewer valid layouts and satisfies a much lower percentage (19\%) of explicitly requested constraints.
Using procedural program logic, PSDL is able to satisfy many explicit constraints on the standard prompt set (83\%), but this comes at the cost of producing far fewer valid layouts (63\%).

\textbf{Ablating concept co-design. }
We also apply this evaluation to the \textit{w/o user guidance} condition (Section~\ref{sec:res_encoding}), which uses the concept implementations initially produced by the LLM, without subsequent user feedback.
As shown in Figure~\ref{fig:validity_adherence}, using co-designed concepts improves both physical validity and prompt adherence.
Together with the concept-level results in Figure~\ref{fig:concept_encoding}, this shows that user feedback not only aligns individual concept behavior with the user's intent, but also helps produce concepts that the LLM can more reliably compose together.

\textbf{Qualitative analysis. }
Figure~\ref{fig:fig_qual_comp} shows qualitative comparisons, along with some annotations on violated constraints.
These comparisons show that \methodname more reliably respects the domain-specific concepts expressed in the prompt.
Holodeck uses low-level constraints that make it difficult to represent higher-level sub-structures (living room conversation areas) or maintain constraints in crowded spaces (keeping a bed in a specific corner).
PSDL produces regularly arranged layouts, but often fails to capture more subtle low-level relationships (arranging sofas into an L-shape).
PSDL also has a catastrophic failure mode where its error correction can over-correct tightly constrained spaces (scattering cutlery off of a table).

\textbf{Human perceptual study. }
To complement our automatic judgments of prompt adherence, we conduct a perceptual study with human raters.
We compare \methodname against PSDL, the strongest baseline on prompt adherence.
Raters see the prompt and two rendered layouts in randomized order, and select which better respects the request or indicate a tie.
In total, we collected 180 responses from 9 participants and found that raters strongly preferred \methodname, choosing its layout 76\% of the time over PSDL generations (excluding tie votes).

\subsubsection{Analysis of Prompt and Program Complexity}
\label{sec:res_complexity}

Across the three comparison domains, prompts in our standard set contain on average $\sim$70 words and $\sim$7 explicitly requested layout constraints.
These prompts are substantially more detailed than the short prompts typically presented in the Holodeck and PSDL evaluations.
Averaging over all domains, LLMs wrote \methodname layout programs that contained $\sim$150 lines of code and, per prompt, instantiated $\sim$21 entities, $\sim$16 \edgenames, $\sim$3 hierarchical \nodenames, and $\sim$2 child concept-graphs.

To study how performance changes with more complex requests, we additionally compare \methodname and PSDL on a \textit{challenge} set of 30 prompts across the three comparison domains.
These prompts average $\sim$220 words and $\sim$18 explicitly requested constraints, resulting in \methodname layout programs that instantiate $\sim$38 objects per generation.
On this set, \methodname maintains a 97\% physically valid generation rate, while PSDL drops from 63\% on the standard prompts to 37\%.
The validity gap between the methods therefore grows from approximately 37 to 60 percentage points as prompt complexity increases.
Despite the additional complexity, \methodname is able to maintain a strong prompt adherence of 78\%, slightly down from 90\% on the standard set.
Appendix~\ref{sec:app_challenge_results} reports full results on this set.

\subsection{Ablation on Concept Composition}
\label{sec:res_ablation}

Even with access to a well-designed library, LLMs must decide how to compose concepts to form a coherent layout.
We ablate our concept-graph intermediate representation, program-repair procedure, and solver machinery used by \methodname.
For these ablations, we follow the procedure described in Section~\ref{sec:res_text_to_layout} to generate layouts across all five domains.
Table~\ref{tab:ablation} reports how often each variant finds a layout that satisfies the generated concept-graph (\textit{solved}), the amount of residual graph error, the number of base validity violations, and solve time in seconds.
Appendix~\ref{sec:app_ablations} details each condition, and the bottom of Figure~\ref{fig:fig_qual_comp} shows qualitative comparisons for selected ablations.

\input{tables/ablation_table}

\textbf{Authoring layout programs. }
The \textit{w/o graph} condition removes the concept-graph structure and requires the LLM to decide how to schedule each concept's synthesis and verification functions directly.
As shown in Figure~\ref{fig:validity_adherence}, this reduces the average valid-generation rate across the three comparison domains down to 30\%, suggesting that it is easier for LLMs to specify which relationships should hold than to search directly for a satisfying layout. 
The \textit{w/o hierarchy} condition allows only a single root concept-graph, so concepts can only reference leaf nodes.
From Table~\ref{tab:ablation}, we observe that removing hierarchy lowers the solve rate from 81\% to 73\%.
The \textit{w/o LLM correction} condition removes the loop that feeds concept-level solver errors back to the LLM for program revision (Section~\ref{sec:program_repair}).
Removing this loop reduces the solve rate from 81\% to 35\% and increases base validity errors from 0.065 to 0.975 violations per layout.

\textbf{Searching for satisfying layouts. }
Finally, we ablate the solver itself (Section~\ref{sec:local_solve}).
\textit{W/o solver repair} removes the verification-driven repair loop, so the solver runs synthesis when concepts initially fail but does not revisit failing concepts.
\textit{W/o verification} runs synthesis blindly, without checking whether a concept is already satisfied.
Both synthesis and verification are needed to reliably search for satisfying layouts: the full solver achieves an 81\% solve rate, compared with 48\% without solver repair and 13\% without verification.

\input{figs/res_fig_concept_evolution}

\subsection{Non-Expert Concept Co-design}
\label{sec:res_user_study}

We next study whether non-expert users can use \methodname to create concepts that match their intent.
We recruit eight participants for a two-part study.
In the first part, participants use \methodname to create concepts from a given target specification.
In the second part, participants propose and create a new concept of their own.
See Appendix~\ref{sec:app_user_study} for the full protocol.

\textbf{Creating concepts. }
In total, participants created 20 concepts across three domains.
In both parts, all participants agreed or strongly agreed that the concept they co-designed matched the intended behavior.
It took participants around 15 minutes on average to co-design a concept.
Participants provided free-form text guidance ($\sim$4.5 per concept) and used the \textit{auto-fix} mode to improve verifiers ($\sim$2.5 applications per concept).

Figure~\ref{fig:concept_evolution} illustrates how two concepts evolved during the user study:
the top row comes from the first part, in which the participant is given a specification, and the bottom from the second, in which the participant proposes their own concept.
In both cases, the LLM's initial implementation differs from the expected behavior and is progressively refined through user guidance.
In the top row, the initial synthesis wedges the bed diagonally into the corner.
The participant corrects this through natural-language guidance in the synthesis view, then notices that the verification function wrongly accepts a nightstand at the foot of the bed and corrects it by labeling validation-set examples and invoking \textit{auto-fix}.
In the bottom row, the participant creates their own \texttt{LookAt} \edgename with a \texttt{preserve\_orientation} mode that keeps each object's original orientation whenever possible.
They specify that neither object should turn, but should instead slide along its wall until the objects face one another.
After aligning the synthesis function with their intent, the participant notices that the verification function still accepts cases in which an object deviates from its preferred orientation, and corrects it through validation-set labeling and \textit{auto-fix}.
By the time each participant approves their concept, synthesis and verification agree on the intended behavior without the participant having authored any code by hand.

After the study, we interviewed participants and observed common trends in their described experiences.
Participants found the interface helpful in defining a concept's behavior, describing it as ``easy to use'', where one participant further commented that the process seemed ``faster than writing pure code, or even writing code with an LLM agent''.
A common thread was that users found the application scenarios beneficial, as being able to easily visualize execution behaviors in different settings ``sped up the design process''.
Another user commented that ``seeing different instantiations'' both ``forced me to think about test cases I wanted'', and moreover, helped them figure out the ``exact behavior I wanted more of''.
Others similarly appreciated how this collaborative co-design process helped them clarify what they wanted: ``it felt like I was forming a concept with the system''; ``we were converging on a specification instead of going back and forth''.

\input{figs/res_fig_glp_qual}

\textbf{Analysis of concept consistency. }
In the first part of the study, users were provided with a concept specification defined by an expert.
As one way to analyze whether the co-designed concepts converged toward expert-aligned semantics, we compare the concepts created by participants against an expert reference built from the same specification.
On held-out test scenarios, participants' initial verifiers (before any guidance) agreed with the expert verifier on only 59\% of expert-valid layouts, while agreeing on 92\% of expert-invalid ones (i.e., they over-reject).
After participants iterated on concept behavior, through free-form feedback and the \textit{auto-fix} procedure, recall rose to 74\% while specificity remained high at 89\%.
This indicates that lightweight participant guidance through \methodname measurably shifts an LLM's first-draft concept towards a consistent semantic interpretation, as proxied by the expert-approved implementation.

%% file: figs/domain_res_fig.tex
\begin{figure*}[t!]
\centering
\includegraphics[width=\textwidth]{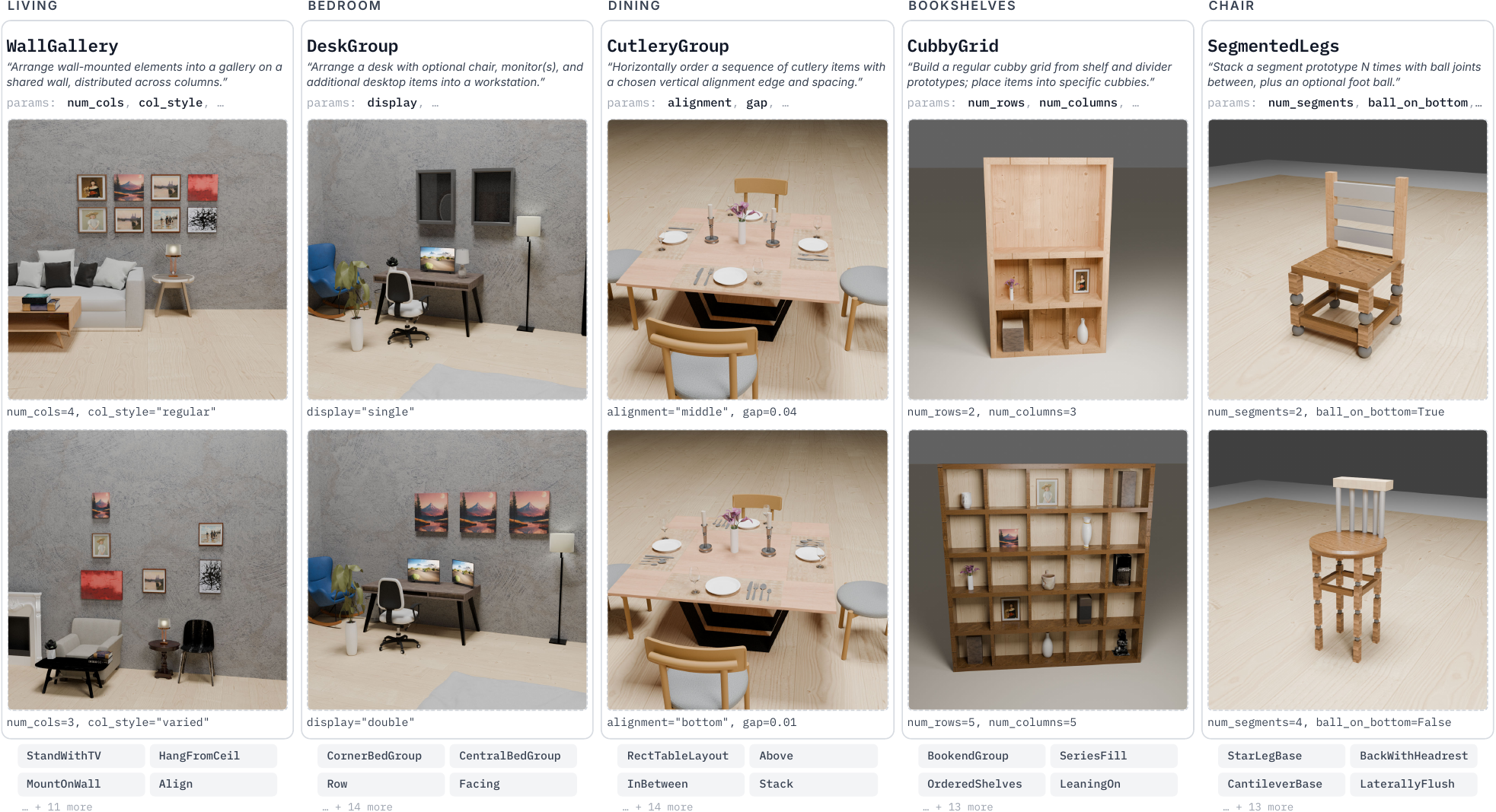}
\caption{Concept libraries co-designed with \methodname. We visualize representative concepts across our evaluation domains. Each domain highlights one concept: we provide a summary of its intent, a subset of its exposed parameters, and show how varying those parameters allows for diverse layout generation.} 
  \label{fig:domain_res}
\end{figure*}

%% file: figs/res_fig_concept_encoding.tex
\begin{figure*}[t!]
\centering
\includegraphics[width=0.93\textwidth]{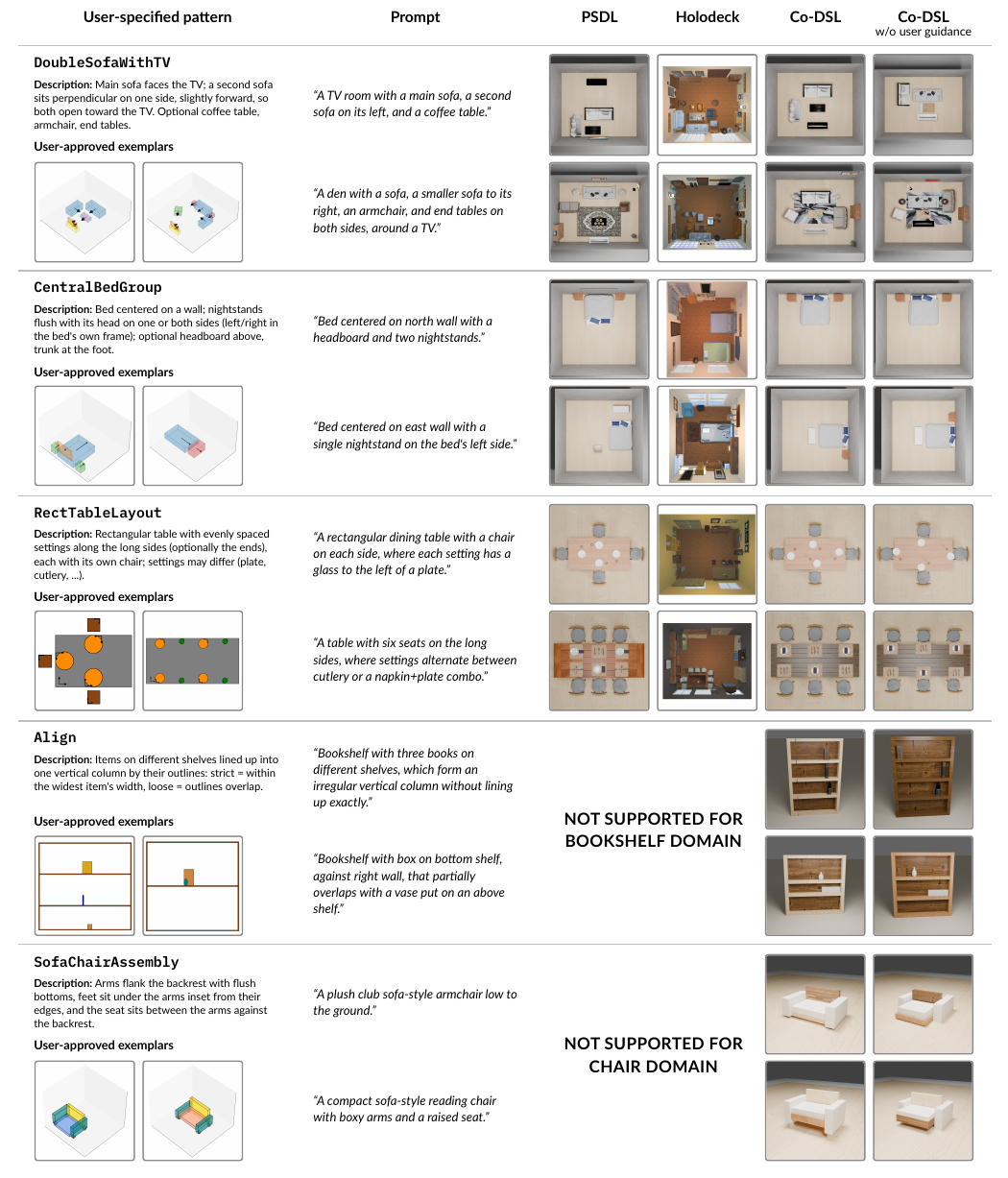}
\caption{
Encoding spatial patterns as reusable concepts.
Each row shows a representative concept from one of our five evaluation domains, including its description and layouts approved by the user during co-design.
For two new prompts requesting the pattern, we compare layouts generated by PSDL, Holodeck, and \methodname.
PSDL and Holodeck receive the user's concept description as part of the prompt, while \methodname uses the co-designed library concept.
The \textit{w/o user guidance} condition instead uses the concept implementation initially produced by the LLM, before the user provides feedback.
PSDL and Holodeck do not support bookshelf arrangements or 3D shape assemblies, which are marked as unsupported.
}
\label{fig:concept_encoding}
\end{figure*}

%% file: figs/res_fig_validity_adherence.tex
\begin{figure*}[t!]
\centering
\includegraphics[width=\textwidth]{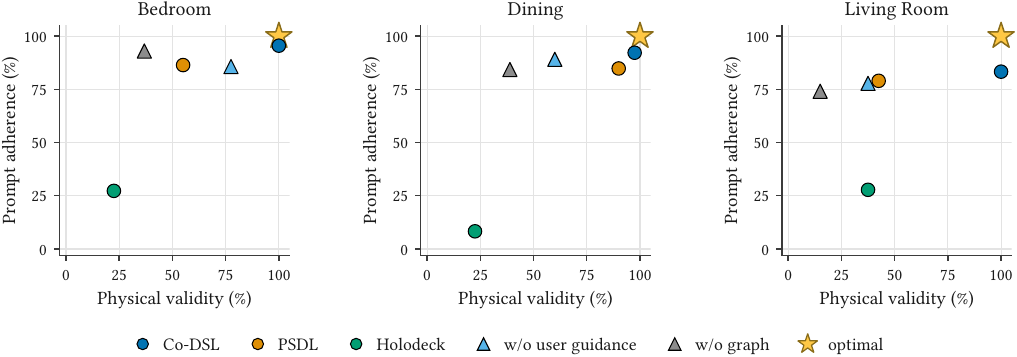}
\caption{
Physical validity versus prompt adherence.
For each method we plot the percentage of generations that pass all base validity checks (X-axis) against the percentage of explicitly requested constraints judged satisfied (Y-axis) on the standard prompts from the bedroom, dining, and living-room domains.
We compare \methodname against recent LLM program synthesis baselines that use fixed operation sets (PSDL, Holodeck) and ablated versions of our method (triangle markers).
The star (top-right corner) marks ideal task performance.
}
\label{fig:validity_adherence}
\end{figure*}

%% file: figs/res_fig_qual_comp.tex
\begin{figure*}[t!]
\centering
\includegraphics[width=.9\textwidth]{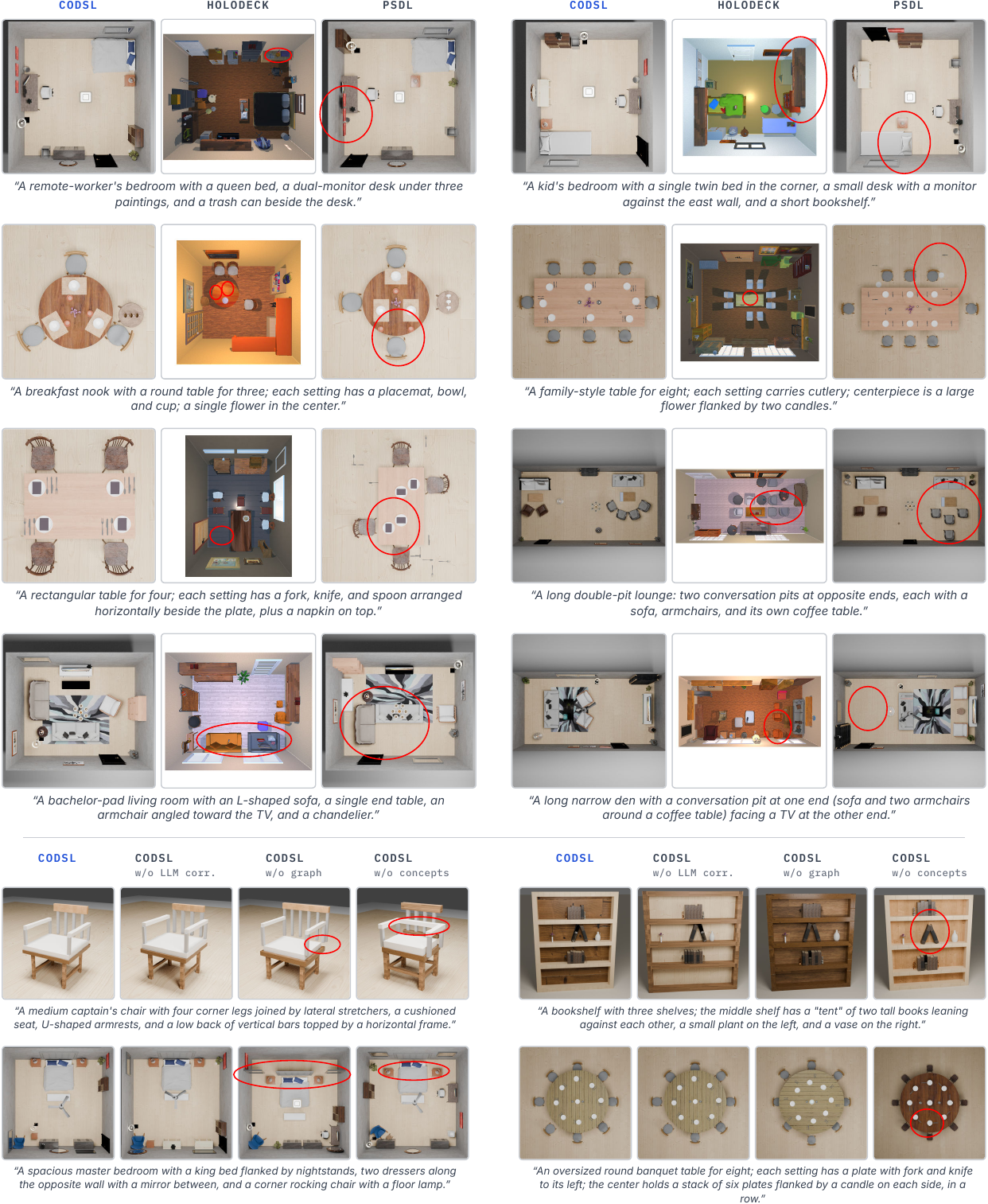}
  \caption{
  Generating layout programs from text.
  We qualitatively compare \methodname against: (\textit{top}) baselines that use LLMs to author programs with fixed operation sets (Holodeck~\cite{yang2024holodeck} and PSDL~\cite{gumin2025proceduralsceneprogramsopenuniverse});
  (\textit{bottom}) ablated versions of \methodname that remove LLM program repair, concept-graphs, and concept libraries.
  We observe \methodname produces more coherent layouts that better respect library concept semantics, adhere better to prompt specifications, and remain physically valid.
  We mark violations with red circles.
  }
  \label{fig:fig_qual_comp}
\end{figure*}

%% file: tables/ablation_table.tex
\begin{table}[t!]
    \centering
    \caption{Ablations on concept composition.  We report how often each variant finds a layout that satisfies the generated concept-graph, how much error is incurred in the best found layout, the number of base validity check violations, and the time taken for each search.
    We describe the ablation conditions in Section~\ref{sec:res_ablation}.
    }
    \footnotesize
    \begin{tabular}{@{}l@{}cccc@{}}
        \toprule
        \textbf{Method}
        & \textbf{Solved\% $\uparrow$}
        & \textbf{Error $\downarrow$}
        & \textbf{Violations $\downarrow$}
        & \textbf{Time $\downarrow$} \\
        \midrule
        \methodname~w/o \textit{hierarchy}               & 73  & 1.02    & 0.320  & 5.9 \\
        \methodname~w/o \textit{LLM correction}          & 35          & 3.52            & 0.975          & 12.2 \\
        \methodname~w/o \textit{solver repair}           & 48          & 1.52            & 0.285          & 13.1 \\
        \methodname~w/o \textit{verification}            & 13          & 3.22            & 0.238          & 4.7 \\
        \methodname                                      & \textbf{81} & \textbf{0.28}   & \textbf{0.065} & \textbf{4.3} \\
        \bottomrule
    \end{tabular}
    \label{tab:ablation}
\end{table}

%% file: figs/res_fig_concept_evolution.tex
\begin{figure*}[t!]
\centering
\includegraphics[width=\textwidth]{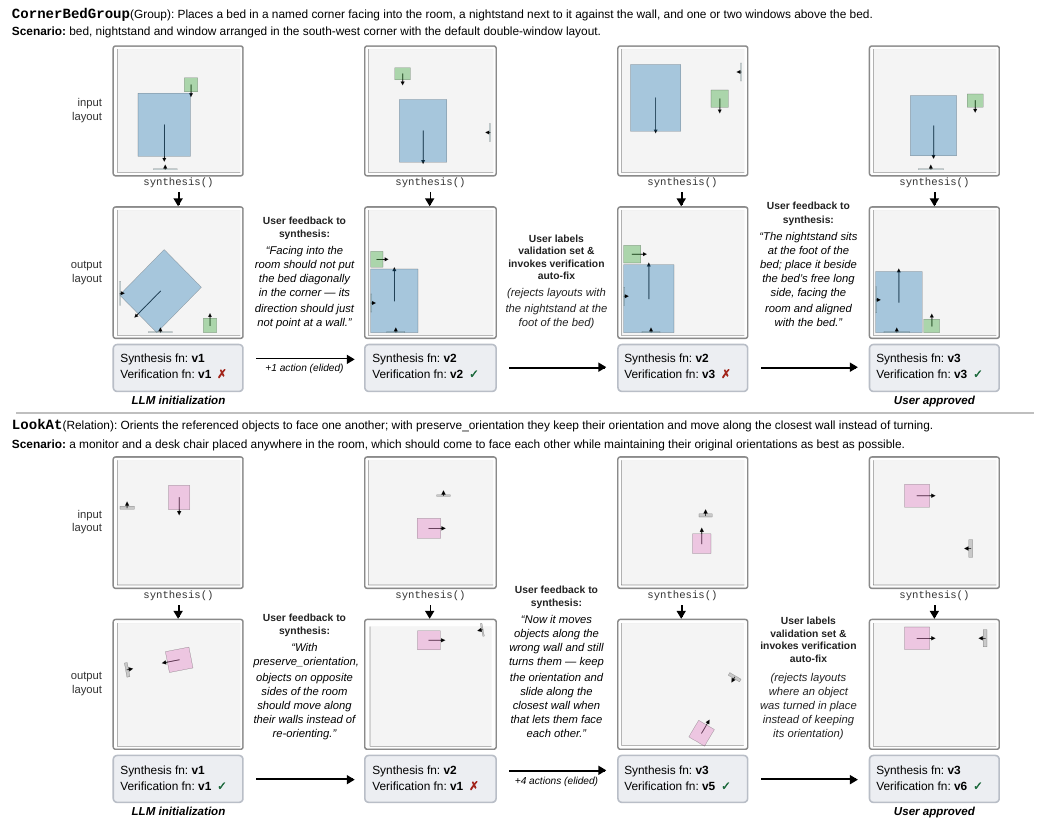}
\caption{
Concept evolution through co-design.
We show the trajectories of two concepts developed by participants in our user study (Section~\ref{sec:res_user_study}).
Each column illustrates the concept's execution behavior on an application scenario:
synthesis transforms the input layout (top) into the output layout (bottom),
while marks next to each verification version record its vote on the synthesized layout.
The leftmost column shows the initial LLM implementation, while the rightmost shows the final version approved by the participant.
Between columns, we summarize the user-provided feedback which drove the LLM to modify the implementation.
}
\label{fig:concept_evolution}
\end{figure*}

%% file: figs/res_fig_glp_qual.tex
\begin{figure*}[t!]
\centering
  \includegraphics[width=\textwidth]{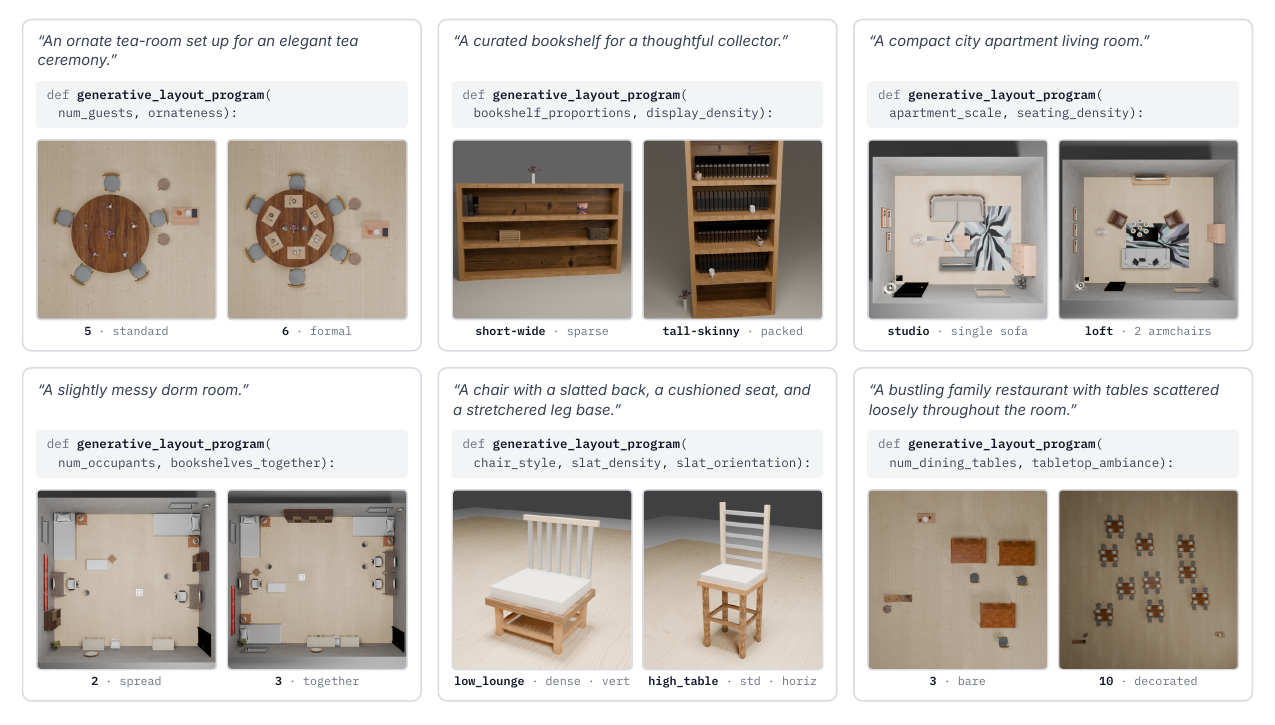}
  \caption{
  Generative layout programs.
  We show examples of LLM authored \metaprogs with \methodname concepts.
  Each layout program exposes a parameterized interface that supports design exploration.
  Changing program parameters produces related layout variations from a single authored program, highlighting the control and diversity afforded by our approach.
  }
  \label{fig:res_fig_glp_qual}
\end{figure*}

%% file: 07-extensions.tex
\section{Extensions}
\label{sec:extensions}

In the previous section, we evaluated \methodname on the task addressed by prior LLM layout systems, producing a single layout from a text prompt.
We find that with access to libraries of \methodname concepts, LLMs are able to realize capabilities that are difficult to achieve with a fixed set of operations.
In this section we demonstrate two such extensions.
First, LLMs can author \metaprogs that expose high-level controls over the layouts they produce (Section~\ref{sec:ext_metaprog}).
Second, users can extend a concept library on demand when a generated layout reveals a conceptual gap (Section~\ref{sec:ext_library}).

\subsection{Generative Layout Programs}
\label{sec:ext_metaprog}

Beyond producing a single layout, a user may want to explore a space of related designs.
Existing LLM program-synthesis systems author programs for individual scenes.
While these programs could in principle be parameterized, reliably mapping user-facing controls to coordinated low-level operations is difficult when the DSL contains only fixed, general-purpose operations.
\methodname makes this problem more tractable: concept-graphs are declarative compositions of high-level concepts, so exposed parameters in a layout program can map directly to concept parameters, concept choices, or graph structure.
We refer to these parameterized functions as \metaprogs.
For example, the right side of Figure~\ref{fig:teaser} shows a program that exposes controls for the number of tables, seats per table, and table arrangement style.

To author a \metaprog, the user describes a scene together with the controls they want to expose, such as the number of diners at a table or the style of its centerpiece.
The LLM returns a layout program whose arguments correspond to these controls.
Given an input parameterization, the program can be executed to instantiate a concept-graph, which is solved with the procedure of Section~\ref{sec:compose_concepts}.
During authoring the LLM also proposes a few example parameterizations, and solver failures on any of them are fed back for repair, so the program is corrected as a whole rather than one layout at a time.
Once authored, the program can be executed for new parameter settings without re-prompting the LLM.

We find that LLMs can use \methodname concepts to author \metaprogs that expose high-level controls over the concept-graphs they instantiate.
We show examples in Figure~\ref{fig:res_fig_glp_qual}, where changing program parameters produces related layout variations from a single authored program.
A low lounge chair with vertical slats can become a high table chair by just changing a few input parameters.
A dorm room can flexibly accommodate different numbers of occupants, and their organizational preferences.
Prior LLM layout program synthesis systems do not offer this kind of control without re-prompting, as they generate individual static scenes rather than reusable controllable layout programs.

We next measure how reliably LLMs can author \metaprogs with \methodname concepts.
We construct an evaluation set of 30 prompts across the bedroom, living room, and dining domains, where each prompt pairs a scene description with two control parameters.
For each prompt an LLM authors two \metaprogs, and we execute each program under three parameterizations.
We measure physical validity and prompt adherence as in Section~\ref{sec:res_text_to_layout}, where parameterizations modify the spatial constraints checked by the LLM judge.
Across the resulting layouts, 93\% of executions are physically valid, and prompt adherence is 84\%.
Further, 83\% of programs produce valid layouts across all evaluated parameterizations.
For comparison, we propose an adaptation of PSDL so that it tries to author parameterized programs in a corresponding setup.
PSDL reaches comparable prompt adherence (84\%), but only 48\% of its executions are physically valid, and only 22\% of its programs remain valid across all evaluated parameterizations.

\input{figs/res_fig_lib_extension}
\subsection{Extending Concept Libraries on Demand}
\label{sec:ext_library}

Our results show that co-designed concept libraries allow LLMs to produce layouts that respect a wide range of spatial patterns.
For any given library, however, a user may still request a pattern that is not well-captured by existing concepts.
In this case the LLM is forced to try to approximate the pattern from ill-suited relations, producing layouts that often violate the intended structure.

Fortunately, \methodname offers a direct way to overcome this failure mode.
If a user notices that an important spatial pattern is poorly modeled by the LLM, they can simply co-design a new concept that captures it.
The new concept can then be added to the library, growing the expressive power available to the LLM, while keeping backwards compatibility with the other library concepts.

Figure~\ref{fig:lib_extension} shows this workflow in three domains.
In each domain, two prompts request a pattern that the initial library does not capture: a reading nook tucked into the corner of a living room, a pair of twin beds with grouped nightstands, and a horseshoe seating arrangement of tables.
With the initial library, the LLM composes the pattern from the concepts it has, and the resulting layouts capture only a part of the requested structure.
After the user co-designs a new concept to capture the pattern, the LLM can author programs from the same prompts to produce layouts that better match the user's intent.
In this way, with \methodname the operations available to the LLM are not fixed in advance, but grow alongside the needs of the user.

%% file: figs/res_fig_lib_extension.tex
\begin{figure}[t!]
\centering
  \includegraphics[width=.9\linewidth]{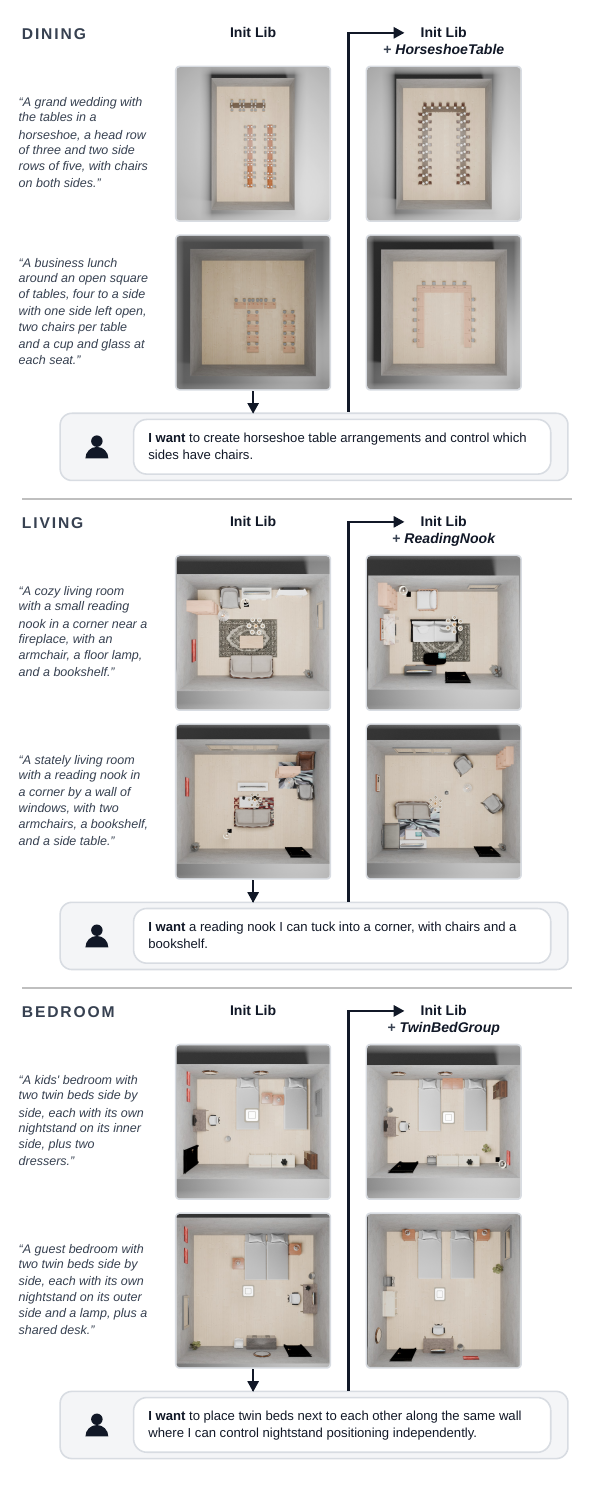}
  \caption{
  Extending a concept library.
  Each domain panel shows two prompts (left) which require a spatial pattern.
  The initial library is not able to represent the pattern (\textit{Init Lib}), but after the user co-designs a new concept, adding it to the library (\textit{+~concept}) produces improved layouts. 
  }
  \label{fig:lib_extension}
\end{figure}

%% file: 08-conclusion.tex
\section{Conclusion}
\label{sec:conclusion}

We introduced \methodname, a framework for human-LLM co-design of programmatic spatial concepts.
Through this process, a user and an LLM work together to jointly develop a library of concepts with agreed-upon semantics, where the user provides feedback on example behavior that guides the LLM in revising the implementation logic.
Given a library, LLMs author layout programs that instantiate concept-graphs, which we can solve to find satisfying layouts.
Across five spatial layout domains, we showed that \methodname concepts and user guidance during co-design are important for reliably capturing domain-relevant spatial patterns.
We found that LLMs can use \methodname concepts to produce layouts that are more often physically valid and better respect requested spatial relationships compared with layouts produced using fixed, general-purpose operation sets.
Our ablations showed that access to concepts alone is not sufficient: the concept-graph representation, solver, and program repair loop are each needed for LLMs to reliably compose them.
We further demonstrated that LLMs can use these concepts to author \metaprogs that expose high-level controls, and that users can extend a concept library on demand when a prompt reveals a missing pattern.
  
\textbf{Limitations and Future Work. }
In our user study, participants reported that providing guidance through our co-design interface allowed them to ``steer the behavior of the concept''.
At the same time, some participants noted that it would be convenient to give feedback through additional forms, such as directly editing a layout or providing sketches that depict an intended pattern.
Exploring how to extend our co-design interface to incorporate additional forms of user input is a natural next step.

To allow users to flexibly design entire libraries of concepts, we make each concept independent of the rest of the library.
This means that concepts can't directly reuse execution behavior to form an abstraction hierarchy, for example, \texttt{Surround} can't call out to \texttt{LookAt}'s logic.
As a partial mitigation, we allow users to include implemented concepts as in-context examples passed into LLM prompts during co-design, but exploring how to leverage concept hierarchy more explicitly could be beneficial.

When LLMs author layout programs with \methodname concept libraries we observe two types of errors:
the LLM may author a layout program that instantiates concept-graphs that cannot be satisfied, or concept-graphs whose solutions do not meet the requirements of the prompt.
We walk through examples of such failure modes in Appendix~\ref{sec:app_failure_modes}.
To mitigate the first issue, we currently rely on LLM repair, which is made possible by the interpretable concept-level errors that verification surfaces.
Though this helps us ultimately produce good layouts, our current repair loop requires interleaving multiple LLM calls with our solver search, and so likely could be made more efficient.
While the second failure mode is not unique to our setting, it would be interesting to explore future work in which an observer (e.g., a user or VLM) inspects the rendered output and proposes corrections back to the LLM.

Finally, the user currently drives all of the library design ideation:
they must think of useful concepts and recognize when a new concept is needed.
It may be possible to aid the user by having the system suggest concept ideas automatically, for example by identifying layout prompts that are poorly covered by the existing library, or by detecting patterns that the LLM frequently attempts to model with low-level operations.

Overall, we view \methodname as an early step toward a promising new paradigm: one in which people and generative AI tools work together to form shared conceptual grounding that can be reused across many future tasks.
While we demonstrate this idea in the context of spatial layout generation, we believe the underlying principle may generalize to other structured reasoning problems.

%% file: appendix.tex
Section~\ref{sec:app_base_domain} describes the per-domain base knowledge for our evaluation domains.
Section~\ref{sec:app_method_details} shares additional method details.
Section~\ref{sec:app_results} reports additional quantitative and qualitative results.
Section~\ref{sec:app_exp_details} gives additional details about the experimental design.
Section~\ref{sec:app_user_study} provides further insights from the non-expert user study.
Section~\ref{sec:app_walkthrough} walks through an LLM authoring sequence of a layout.
Finally, in Section~\ref{sec:app_library} we provide an example implementation of one of our co-designed concepts.

\input{appendix/app_A_base_domain}

\input{appendix/app_B_method_details}

\input{appendix/app_C_additional_results}
\input{appendix/app_D_experiment_details}

\input{appendix/app_E_user_study}

\input{appendix/app_F_authoring_walkthrough}

\input{appendix/app_G_library}

%% file: appendix/app_A_base_domain.tex
\section{Base Domain Knowledge}
\label{sec:app_base_domain}

\begin{figure*}[t!]
\centering
\includegraphics[width=\textwidth]{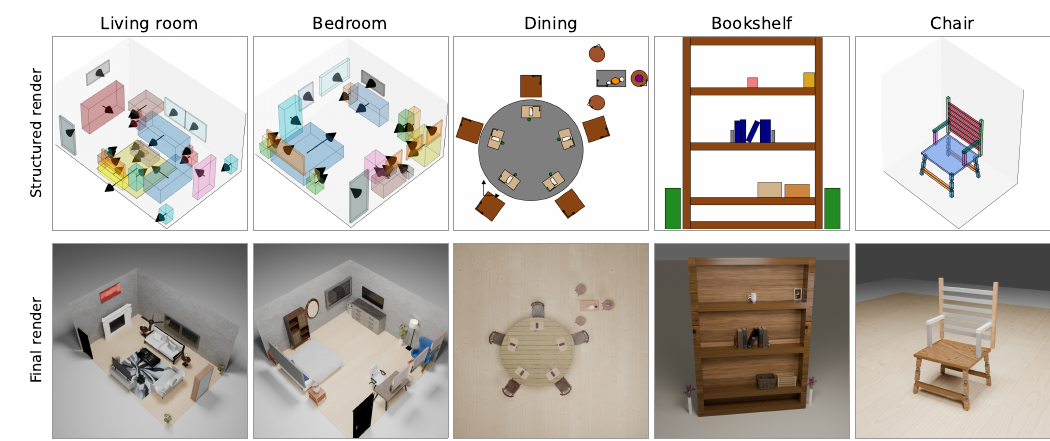}
\caption{Structured layout domains.
For each domain, we show an example layout rendered by the domain's structured logic (top) and as a 3D Blender scene using our final rendering pipeline (bottom).}
\label{fig:app_structured_vs_final}
\end{figure*}
\methodname requires shared per-domain information that all concepts can rely on and that we expose to the LLM during layout program authoring.
This base knowledge supports concept implementation and execution, but does not encode the recurring spatial patterns that a user might care about.          
We organize this into the four fields introduced in Section~\ref{sec:domain_reqs}: \textbf{spatial bounds and frames}, an \textbf{entity model}, \textbf{scene initialization}, and \textbf{base validity checks}.
Each domain also provides a \textbf{rendering procedure}, which is used only after concept-graphs have been solved, to convert structured layouts into images (Section~\ref{sec:app_rendering}).
As detailed below, we find much of this infrastructure can be reused across related layout domains.

\subsection{Family: 3D Rooms}

\textit{Living rooms} and \textit{bedrooms} share the same family base.
The only per-domain difference is the atomic objects available to each domain.

\textit{Spatial bounds and frames.}
A rectangular 3D room with width, height, and depth, where the floor is the bottom plane, the ceiling is the top, and four vertical walls bound the remaining sides.
The X-axis is left-to-right (negative-positive), the Y-axis is bottom-to-top (negative-positive), and the Z-axis is back-to-front (negative-positive).

\textit{Entity Model.}
Each object is an upright 3D oriented bounding box: center position, three side lengths, and a yaw about the vertical Y axis.
A \textit{placement type} tag indicates how the object relates to room surfaces: \textit{regular} (rests on floor or another regular object), \textit{floor-anchored} (e.g.\ a rug), \textit{wall-anchored} (e.g.\ a painting), or \textit{ceiling-anchored} (e.g.\ a fan).
Concepts call into the base for geometric queries (e.g. extent, orientation) and transformations (e.g. translation, planar rotation), applied to either objects or groups of objects.

\textit{Scene initialization.}
Regular and floor-anchored objects start at a uniform random floor position, wall-anchored objects start on a randomly chosen wall facing inward, and ceiling-anchored objects start attached to the ceiling.

\textit{Base validity checks.}
There are three checks:
(a) every object lies inside the room;
(b) objects do not penetrate one another, except for floor-anchored objects beneath regular objects;
(c) every object has a path to the floor or the wall, where regular objects can rest on the floor through a chain of supporting objects.

\subsection{Family: 2D layered arrangements}

\textit{Dining arrangements} (top-down) and \textit{bookshelves} (front elevation) share the same planar-style scene base.
The two in-plane axes are interpreted by the domain, and the third spatial dimension is handled abstractly by an integer \textit{layering parameter} rather than by an explicit coordinate.
The two domains differ in how they interpret this parameter and in a small set of extensions on top.

\textit{Spatial bounds and frames.}
A bounded 2D rectangle.
The domain specifies its orientation (top-down vs.\ elevation), whether one axis is gravity-aligned, and whether implicit reference geometry is present.

\textit{Entity Model.}
Each object is an oriented 2D bounding box with a center position, two side lengths, and an in-plane orientation, and carries an integer layering parameter whose meaning is specified by the domain (physical stacking, or front to back placement).
Geometric queries return planar extent, corners, and per-group layering extents.
Transformations cover planar translation, in-plane rotation, and layering manipulation.

\textit{Scene initialization.}
Each object is initialized at a position inside the scene that respects its category, using a procedure provided by the domain.
Some items are placed randomly.
Items with known supports are sampled so that they initially rest on a supporting object. 

\textit{Base validity checks.}
There are two checks shared by both domains:
(a) every object lies inside the scene;
(b) the projections of objects in the same layer do not overlap.
There are additional checks for each sub-domain.

\paragraph{Dining arrangements.}
Top-down view of a dining arrangement.
The layering parameter models \textit{physical stacking}: level $0$ is on the ground, higher levels rest on objects below, and validity adds a containment rule that every object above level $0$ must lie inside the projection of some object on the level below. 

\paragraph{Bookshelves.}
Front elevation view of a bookshelf, where gravity points along the negative vertical direction and the ground is the bottom edge of the scene. 
The layering parameter is reinterpreted as front to back object placement. 
Each scene has an implicit frame made of four boards, which are created automatically and parameterized by width, height, and board thickness.
Bookshelves add additional validity checks: 
non-structural items must rest on a support (e.g. a frame), 
tilted items need a prop in the same layer on the leaning side, 
and structural members must lie inside the frame and chain back to it through structural anchors (no floating shelves). 

\subsection{Family: 3D shape assemblies}

\textit{Spatial bounds and frames.}
A bounded 3D envelope sized to the expected shape.
The X-axis is left-to-right (negative-positive), the Y-axis is bottom-to-top (negative-positive), and the Z-axis is back-to-front (negative-positive).

\textit{Entity Model.}
Each part is a 3D oriented primitive: center position, three side lengths, and three direction vectors, along with a \textit{proxy shape} tag (box, sphere, or cylinder). 
Each part also has a semantic role consumed by validators and concept verifiers (e.g., leg or seat parts for chairs).
Geometric queries return extent, corners, and orientation, while transformations allow translation and full 3D rotation about a pivot axis.

\textit{Scene initialization.}
Parts are positioned at random horizontal positions within the envelope, with vertical positions set so the part's bottom rests on the floor.

\textit{Base validity checks.}
There are two checks:
(a) every part lies inside the envelope;
(b) every part chains to the floor through contact, so no part may float.

\subsection{Final 3D Rendering Pipeline}
\label{sec:app_rendering}

The native rendering procedure of each domain draws a layout as a \emph{structured} view (Figure~\ref{fig:app_structured_vs_final}, top).
For a more realistic rendering pipeline, as used in the qualitative results throughout the main paper, we convert each structured layout into a final 3D scene rendered in Blender~\cite{blender}.
For the non-chair domains we build an asset library by downloading BlenderKit~\cite{blenderkit} assets for each category, normalizing them to a common format, and labeling each asset with its semantic category and front-facing direction.
At render time, an asset is matched to each bounding box by category and dimension.

Domains in the 3D rooms family place retrieved assets directly.
Domains in the 2D layered arrangements family first lift the layout into 3D using the layering parameter, which gives stacking levels for dining arrangements and front to back placement inside a frame of fixed depth for bookshelves, and then retrieve assets in the same way.
The chair domain works differently, as we render each part directly as its proxy primitive with a simple material scheme (wood for legs, fabric for cushions, metal for pedestals).
Figure~\ref{fig:app_structured_vs_final} shows one example from each domain using both the domain's structured rendering procedure and the final rendering script.

%% file: appendix/app_B_method_details.tex
\section{Additional Method Details}
\label{sec:app_method_details}

\subsection{Solver Details}
\label{sec:app_solver}

Algorithm~\ref{algo:solve_cg} in Section~\ref{sec:compose_concepts} provides pseudocode for our solving procedure.
We discuss additional implementation details here.

For our main experiments, we give the outer solve loop a time budget of 15 seconds and stop early if it finds a layout with zero error.
In each iteration of this outer loop, we execute the layout program to instantiate a concept-graph and then solve it.
Because different executions of the same layout program can instantiate structurally different concept-graphs, repeated iterations give the solver alternative graphs to search over.

Any given concept-graph may contain child concept-graphs as nodes.
We solve each child first and insert its solved sub-scene into the parent as a single node.
The parent's concepts can then transform this sub-scene as a unit, but cannot change its internal arrangement.
In this way, local solving at a parent graph begins only after every child has been replaced by a solved sub-scene.

\textbf{Scope of Concept Execution.}
A concept's \textit{Synthesis()} and \textit{Verification()} functions have access to the same inputs: the concept parameters, the bounds of the coordinate space, and the entities referenced by the concept.
Neither function can inspect other entities in the scene.
Synthesis can modify the referenced entities, and \nodename concepts may additionally create or copy entities.
Verification checks whether the referenced entities satisfy the concept without modifying them.
Consequently, a synthesis update may pass its own verification while introducing a collision with an unreferenced object or violating another concept's requirements.
The solver evaluates these interactions by combining the verification of each concept with base validity checks over the whole scene.

After the synthesis pass we compute a local graph error by aggregating base validity failures, unsatisfied concept verifications, and errors propagated from any child concept-graphs.
Base validity failures are weighted at $2.0$ each, failed concept verifications at $1.0$ each, and concept synthesis calls that cannot be scheduled because of conflicting dependencies at $0.5$ each.
Base validity failures receive the greatest weight because they indicate physical errors that obstruct further constraint satisfaction.
Unscheduled synthesis calls receive the smallest weight because the concept may still be repaired downstream.
If the error is nonzero, we enter a verification-driven repair loop that revisits failing concepts and reapplies their \textit{Synthesis()} function, keeping updates only when they reduce the total error.

We maintain a cache for each concept-graph, so that once a sub-scene has been solved, later attempts to solve the larger graph can reuse it directly.
At the end of each call we update the cache if the current scene improves on the cached one.

\subsection{Concept-Graphs in Relation to Prior Work}
\label{sec:app_cg_comparison}

As discussed in Section~\ref{sec:rel_work}, our concept-graph framework follows from a long line of structured representations for modeling and representing visual data.
Factor graphs, relation graphs, and constraint graphs similarly state constraints over scene entities and then solve for a layout that satisfies them~\cite{LARJ,fisher2011,Para2021ConstraintGraphs}.
The main difference in our setting is that the user and the LLM co-design the operation set, instead of using a set of constraints that are fixed in advance.
The intermediate representation and solver must therefore support concepts with a wide range of possible behaviors.

There are also two mechanical differences in the representation that differentiate it from prior approaches.
First, a \nodename concept can collect several entities into a reusable group that later concepts treat as one unit.
This directly supports hierarchical composition.
Second, \nodename concepts can add entities during execution, so the graph constructs the scene from prototypes rather than only positioning a set of fixed leaf-entities.

Because each concept exposes dual execution modes, synthesis and verification, a concept-graph simultaneously defines a proposal mechanism and a validity check for layouts.
This allows \methodname to generate layouts using a relatively simple proposal-and-verification strategy.
Prior approaches such as Yeh et al.~\cite{LARJ} instead require specialized inference procedures designed for probabilistic constraint systems.
In contrast, \methodname places more structure in the representation itself and can use a more lightweight solver.

Farooki et al.~\cite{farooki} introduce another related representation based on design templates associated with functional zones.
Their templates define sub-areas of a layout, and users select placement options from a limited set of rule types for each zone.
This provides a top-down specialization of the layout structure and involves user choices during generation.
In contrast, \methodname adopts a more bottom-up specialization process: the user and the LLM instead co-design concepts that capture recurring layout patterns.
The LLM later composes these concepts into concept-graphs without requiring further user interaction during layout generation.

\subsection{Ablation Conditions}
\label{sec:app_ablations}

We introduce ablations of \methodname to examine the design decisions in our system.
Each ablation removes part of the concept library, the concept-graph representation, the correction loop of Section~\ref{sec:program_repair}, or the solver in Algorithm~\ref{algo:solve_cg}.
We describe how each condition differs from our default settings below:

\textbf{\methodname~w/o \textit{user guidance}} replaces the co-designed library with the initial concept implementations produced by the LLM, before the user provides guidance.
Concept interfaces are fixed before this refinement, so the same layout programs run against both library versions.
The solver and all other algorithmic components are unchanged.

\textbf{\methodname~w/o \textit{graph}} gives the LLM the same concept library but requires it to call the concepts' synthesis and verification functions directly and choose their execution order.
This condition does not construct a concept-graph or use the solver search.

\textbf{\methodname~w/o \textit{concepts}} gives the LLM only the base domain knowledge and no library concepts.
The LLM writes object positions directly.

\textbf{\methodname~w/o \textit{hierarchy}} permits only a single root concept-graph.
Concepts may reference only leaf entity nodes, so the program cannot use child concept-graphs or pass \nodename nodes into other concepts.

\textbf{\methodname~w/o \textit{LLM correction}} uses the LLM's initial layout program as $P$ in Algorithm~\ref{algo:solve_cg}, before any rounds of the layout program correction loop are run.
The solver and library are unchanged.

\textbf{\methodname~w/o \textit{solver repair}} runs \textproc{SolveLocal} without the inner verification-driven repair loop of Algorithm~\ref{algo:solve_cg}, so each concept's \textit{Synthesis()} is applied at most once per attempt, and verification failures are not used to drive further updates.

\textbf{\methodname~w/o \textit{verification}} removes concept-specific verification entirely, so the solver applies each concept's \textit{Synthesis()} without first checking its \textit{Verification()}, and the graph error counts only base validity failures.
This isolates the contribution of the concept-specific verification functions.

%% file: appendix/app_C_additional_results.tex
\section{Additional Results}
\label{sec:app_results}

\begin{figure*}[!tp]
\centering
\includegraphics[width=\textwidth]{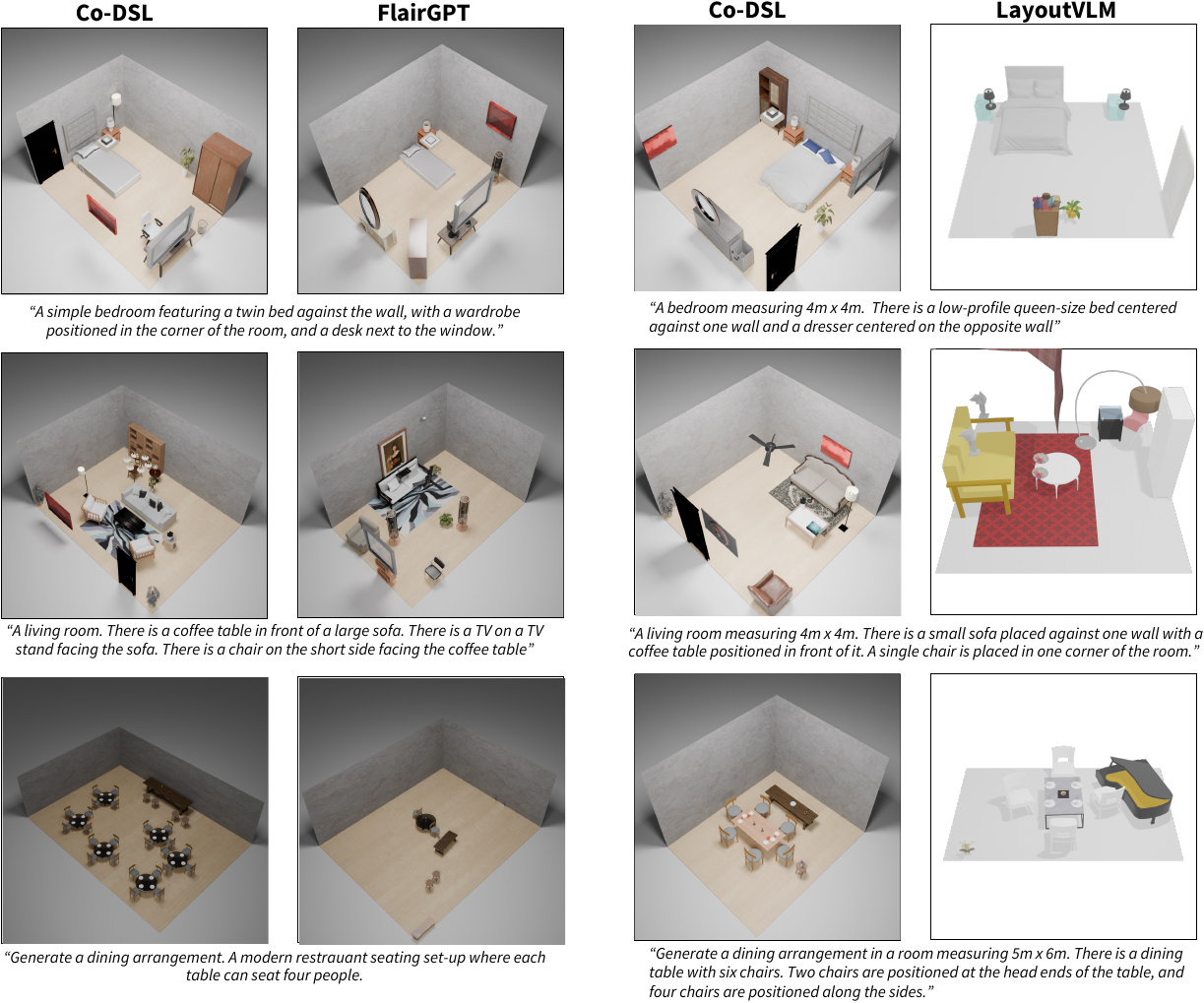}
\caption{Qualitative comparison of \methodname against FlairGPT~\cite{littlefair2025flairgpt} and LayoutVLM~\cite{sun2025layoutvlmdifferentiableoptimization3d} on bedrooms, living rooms, and dining arrangements.
Each row shows a prompt and the layout produced by each method.}
\label{fig:supp_flair_layoutvlm}
\end{figure*}

\begin{figure*}[!tp]
\centering
\includegraphics[width=\textwidth]{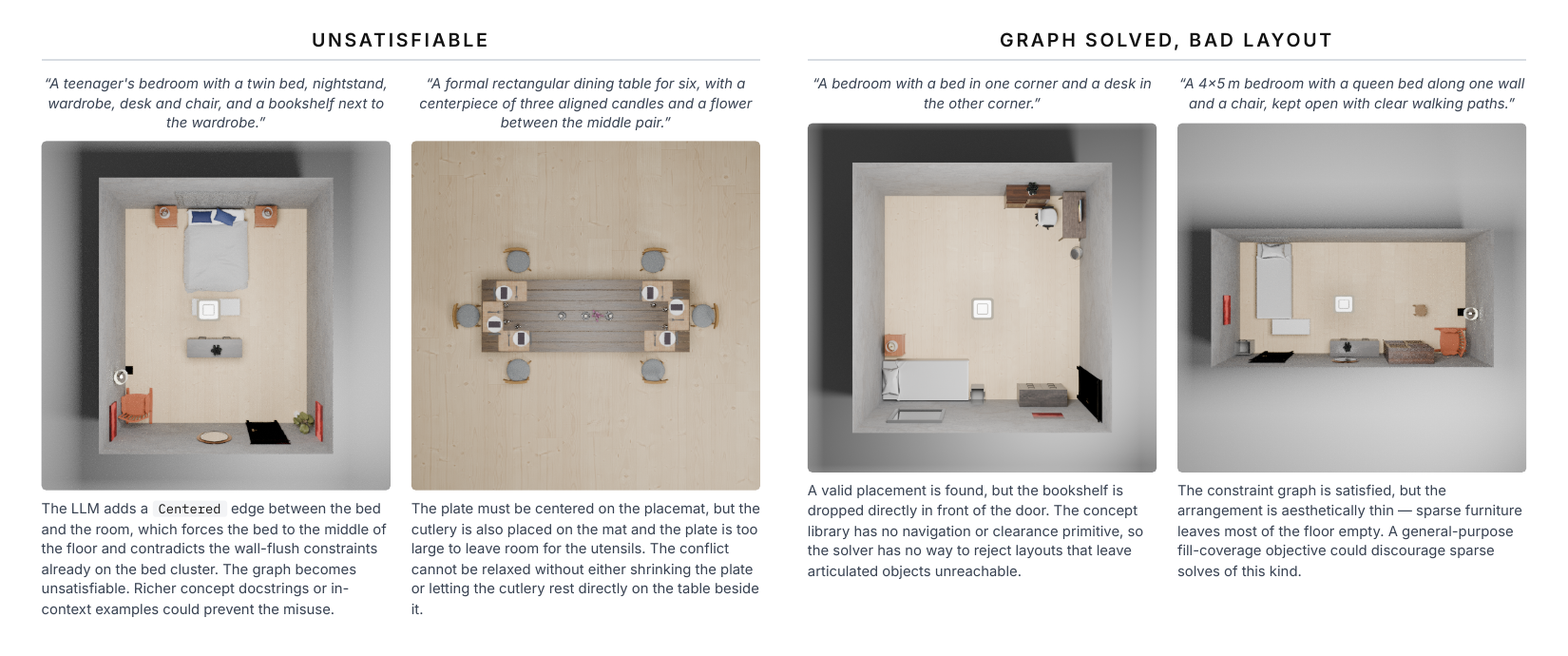}
\caption{Failure modes of scenes generated by \methodname.
\textit{Unsatisfiable} (left): the LLM writes a program whose concept-graphs admit no valid placement.
\textit{Graph solved, bad layout} (right): the solver finds a satisfying layout, but the scene violates implicit expectations that no constraint in the graph captures.}
\label{fig:supp_failure_modes}
\end{figure*}

\subsection{Per-Domain Base Validity}
\label{sec:app_validity_table}

\input{appendix/data/validity_table}

Table~\ref{tab:domain_solve_comp} reports base validity for each method and domain.
Each cell gives the average number of base validity violations and the percentage of layouts that pass all base checks, computed over completed generations.
A dash indicates that a method does not support the domain.
Every method completes all of its generations except for two ablations: \textit{w/o graph} does not complete a generation for one bedroom, two dining, and one bookshelf prompt, and \textit{w/o concepts} does not complete one living-room prompt.

\subsection{Per-Domain Prompt Adherence}
\label{sec:app_adherence_table}

\begin{table}[t!]
\centering
\footnotesize
\caption{Prompt adherence.
We report the percent of explicitly requested prompt constraints judged satisfied (Section~\ref{sec:app_metrics}).}
\label{tab:app_adherence}
\begin{tabular}{lcccc}
\toprule
\textbf{Method} & \textbf{Living} & \textbf{Bedroom} & \textbf{Dining} & \textbf{All} \\
\midrule
Holodeck                 & 27.8 & 27.3 &  8.3 & 19.4 \\
PSDL                     & 79.0 & 86.4 & 84.8 & 83.5 \\
\midrule
\methodname w/o user guidance   & 77.8 & 85.7 & 89.1 & 84.8 \\
\methodname w/o graph           & 74.1 & 93.0 & 84.3 & 83.5 \\
\methodname w/o concepts        & 90.8 & 90.9 & 93.9 & 92.2 \\
\midrule
\methodname                     & 83.3 & 95.5 & 92.2 & 90.5 \\
\bottomrule
\end{tabular}
\end{table}

Table~\ref{tab:app_adherence} reports prompt adherence results split by domain.
We use the same judge and protocol for every method and condition, as described in Section~\ref{sec:app_metrics}.
We judge every completed generation regardless of its physical validity, so an invalid layout may still receive a high adherence score.
The pooled values correspond to the results discussed in Section~\ref{sec:res_text_to_layout}.
\methodname satisfies 90.5\% of requested constraints, compared with 83.5\% for PSDL and 19.4\% for Holodeck.

The \textit{w/o concepts} condition achieves the highest overall prompt adherence score.
This comes at a significant cost: the same condition passes the base validity checks in only 26.3\% of living rooms, 15.0\% of bedrooms, and 35.0\% of dining arrangements.
Without library concepts, the LLM can write coordinates that satisfy many requested relations but often produces physically invalid layouts.
As depicted in Figure~\ref{fig:validity_adherence}, the results from these two tables should therefore be considered together.

\subsection{Challenge Prompt Set Results}
\label{sec:app_challenge_results}

We additionally compare \methodname and PSDL on the challenge prompt set described in Section~\ref{sec:app_prompt_sets}.
We generate one layout per prompt and method, and both methods complete all 30 generations.
We evaluate adherence using the same protocol as for the standard set.

\methodname produces valid layouts for 29 of 30 prompts, or 96.7\%.
The one invalid layout contains a single out-of-bounds object in a living room.
PSDL produces valid layouts for 11 of 30 prompts, or 36.7\%: three bedrooms, six dining arrangements, and two living rooms.
\methodname satisfies 77.5\% of the requested constraints, with per-domain scores of 66.7\% for bedrooms, 86.5\% for dining arrangements, and 77.9\% for living rooms.
PSDL satisfies 83.9\% overall, with corresponding per-domain scores of 86.9\%, 73.0\%, and 89.6\%.
PSDL therefore satisfies a somewhat higher fraction of the requested constraints on this set, but this comes at the cost of failing to maintain physical validity.
This pattern is to be expected for procedural methods that can directly encode the specific relations expressed in the prompt.
As prompts become more demanding and add more conflicting semantic constraints, programs that use low-level relations to jointly satisfy all of them tend to fail, which for PSDL manifested as violations of base physical constraints.

\subsection{Human Perceptual Study}
\label{sec:app_perceptual_results}

\begin{table}[t!]
\centering
\footnotesize
\caption{Human perceptual study.
For each domain, we report the percentage of responses in which raters preferred the \methodname layout, preferred the PSDL layout, or expressed no preference, with three raters and 20 pairs per rater in each domain (60 responses per domain).}
\label{tab:app_pref}
\setlength{\tabcolsep}{4pt}
\begin{tabular}{lccc}
\toprule
\textbf{Domain} & \textbf{\methodname} & \textbf{PSDL} & \textbf{No preference} \\
\midrule
Living  & 50.0 & 33.3 & 16.7 \\
Bedroom & 76.7 & 11.7 & 11.7 \\
Dining  & 58.3 & 13.3 & 28.3 \\
\bottomrule
\end{tabular}
\end{table}

We compare \methodname against PSDL, the strongest baseline on prompt adherence.
Because both methods arrange the same set of objects for each prompt (Section~\ref{sec:app_comparisons}), the two layouts in a pair differ only in their arrangement.
Nine university students each judged 20 pairs of layouts sampled from the standard prompt set, with three raters per domain, for 180 responses in total.
For each pair, the rater saw the layout prompt and the two rendered layouts in random order and answered the question ``Which generation better respects the requested layout?'', with the option of declaring a tie.
Table~\ref{tab:app_pref} reports the per-domain breakdown of the responses.
Raters preferred the \methodname layout in 61.7\% of responses and the PSDL layout in 19.4\%, with 18.9\% ties; excluding ties, \methodname was preferred 76.0\% of the time.
The preference is strongest for bedrooms and dining arrangements and weakest for living rooms, which mirrors the per-domain adherence gap between the two methods in Table~\ref{tab:app_adherence}.

\subsection{Concept-Adherence Expert Analysis}
\label{sec:app_concept_adherence}

\begin{table}[t!]
\centering
\footnotesize
\caption{Concept-adherence expert analysis.
For each concept and method, we report the number of generated layouts that the expert judged to respect the concept's semantics.}
\label{tab:app_concept_adherence}
\setlength{\tabcolsep}{3pt}
\begin{tabular}{llcccc}
\toprule
\textbf{Domain} & \textbf{Concept} & \textbf{\methodname} & \textbf{w/o guidance} & \textbf{PSDL} & \textbf{Holodeck} \\
\midrule
Bedroom & \texttt{CornerBedGroup}     & 2/2 & 0/2 & 1/2 & 0/2 \\
Bedroom & \texttt{CentralBedGroup}    & 2/2 & 1/2 & 0/2 & 0/2 \\
Bedroom & \texttt{DeskGroup}          & 2/2 & 0/2 & 0/2 & 0/2 \\
Dining  & \texttt{RectTableLayout}    & 2/2 & 0/2 & 0/2 & 0/2 \\
Dining  & \texttt{CircularTableLayout}& 2/2 & 2/2 & 0/2 & 0/2 \\
Dining  & \texttt{Surround}           & 2/2 & 0/2 & 1/2 & 0/2 \\
Living  & \texttt{DoubleSofaWithTV}   & 2/2 & 0/2 & 0/2 & 0/2 \\
Living  & \texttt{ConversationPit}    & 2/2 & 2/2 & 0/2 & 0/2 \\
Living  & \texttt{Facing}             & 2/2 & 1/2 & 2/2 & 1/2 \\
\midrule
\multicolumn{2}{l}{Total}    & 18/18 & 6/18 & 4/18 & 1/18 \\
\bottomrule
\end{tabular}
\end{table}

This analysis uses three concepts from each of the three domains supported by PSDL and Holodeck and two layout prompts per concept, giving 18 layouts per method.
For each layout, the expert records a binary judgment of whether it respects the user-defined concept semantics and provides a short text justification.
Reasons for negative judgments include a nightstand being unreachable from the head of a bed, chairs not facing toward the center of an arrangement, or missing place settings.

\subsection{Generative Layout Program Results}
\label{sec:app_glp_results}

\begin{table}[t!]
\centering
\footnotesize
\caption{Generative layout programs by domain.
`Valid executions' is the percentage of completed executions with zero base validity violations.
`Adherence' is the percentage of object-based and control constraints judged satisfied.
`Valid across range' is the percentage of authored programs for which every execution is valid.}
\label{tab:app_glp}
\setlength{\tabcolsep}{3pt}
\begin{tabular}{llccc}
\toprule
\textbf{Domain} & \textbf{Method} & \textbf{Valid exec.\ (\%)} & \textbf{Adherence (\%)} & \textbf{Valid across range (\%)} \\
\midrule
\multirow{2}{*}{Living}  & PSDL        & 36.7 & 84.5 & 20.0 \\
                         & \methodname & 96.1 & 80.2 & 84.2 \\
\midrule
\multirow{2}{*}{Bedroom} & PSDL        & 70.0 & 84.7 & 35.0 \\
                         & \methodname & 99.2 & 81.4 & 90.0 \\
\midrule
\multirow{2}{*}{Dining}  & PSDL        & 36.2 & 82.1 & 10.0 \\
                         & \methodname & 85.0 & 86.8 & 75.0 \\
\midrule
\multirow{2}{*}{All}     & PSDL        & 47.6 & 83.5 & 21.7 \\
                         & \methodname & 93.4 & 83.5 & 83.1 \\
\bottomrule
\end{tabular}
\end{table}

Table~\ref{tab:app_glp} provides a per-domain breakdown of the results from Section~\ref{sec:ext_metaprog}, with the setup described in Section~\ref{sec:app_extensions_setup}.
Across domains, 93.4\% of \methodname executions are valid and 83.5\% of requested constraints are satisfied.
For \methodname, 83\% of programs produce valid layouts across their evaluated parameter range.
PSDL achieves the same prompt-adherence score of 83.5\%, but only 47.6\% of its executions are valid and only 22\% of its programs remain valid across their range.

\subsection{Additional Qualitative Comparisons}
\label{sec:app_qual_comp}

In the main paper, we compare against PSDL and Holodeck, state-of-the-art text-to-scene generation methods that use LLM program synthesis.
As discussed in Section~\ref{sec:rel_work}, other methods can also perform this task.
Here, we show qualitative comparisons against two additional systems.

We consider FlairGPT~\cite{littlefair2025flairgpt}, which guides an LLM through a multi-step prompting process to incrementally add domain-general constraints that are then handed to a solver, and LayoutVLM~\cite{sun2025layoutvlmdifferentiableoptimization3d}, which uses a vision-language model to propose initial object placements and local constraints that are then optimized by a constraint solver.
We also attempted to run I-Design~\cite{idesign}, but found that roughly half of its LLM calls would hang indefinitely; the layouts we were able to obtain were of similar quality to LayoutVLM, so we omit it from the figure.

Figure~\ref{fig:supp_flair_layoutvlm} shows example results.
In these examples, both methods struggle with compositional scenes that require many relationships between objects.
As complexity grows, the number of low-level constraints also grows, and global consistency becomes difficult to maintain.
The dining arrangements demonstrate this problem: multiple settings must each preserve a stable local structure while being tiled into a larger pattern.
\methodname instead represents such scenes more compactly through hierarchical concept-graphs.
Common sub-scenes are abstracted into reusable concepts (e.g., a place setting or a desk group), and larger structures are composed by arranging those sub-scenes as single units.
\methodname uses an initial LLM call and up to two correction calls to author a layout program, followed by bounded solver execution.
FlairGPT relies on multi-step prompting loops of roughly 20 LLM calls that can take over 30 minutes to converge even on smaller scenes.
In our runs, FlairGPT timed out after one hour of constraint solving in 15 of 72 trials, with failures concentrated on dining arrangements.

\subsection{Failure Modes}
\label{sec:app_failure_modes}

Figure~\ref{fig:supp_failure_modes} depicts two typical failure modes when LLMs author \methodname layout programs.
We describe these failures here and discuss possible future fixes in Section~\ref{sec:conclusion}.

\textit{Unsatisfiable (left).}
The LLM writes a program whose concept-graphs admit no valid placement.
In the bedroom example, the program adds a \texttt{Centered} edge between the bed and the room, which forces the bed to the middle of the floor and contradicts the constraints that already place the bed group flush against a wall.
In the dining example, the plate is required to be centered on the placemat, but the cutlery is also placed on the mat and the plate is too large to leave room for the utensils.

\textit{Graph solved, bad layout (right).}
The solver finds a satisfying layout, but the scene violates implicit expectations that no constraint in the graph captures.
In one case, a bookshelf is placed directly in front of the only door because the concept library has no concept for navigation or clearance.
In another case, the graph is satisfied but the bedroom is sparse, leaving the floor mostly empty because no objective rewards spatial coverage.

%% file: appendix/data/validity_table.tex
\begin{table}[t!]
    \centering
    \footnotesize
    \caption{Base validity of layouts generated from LLM authored layout programs. Each entry gives the average number of validity-check violations per layout ($\downarrow$) followed by the percentage of layouts that are valid ($\uparrow$). Dashes indicate that the method does not produce layouts for that domain.}
    \label{tab:domain_solve_comp}
    \setlength{\tabcolsep}{2.5pt}
    \resizebox{\linewidth}{!}{%
    \begin{tabular}{lccccc}
        \toprule
        \textbf{Method}              & \textbf{Living}     & \textbf{Bedroom}     & \textbf{Dining}     & \textbf{Bookshelf} & \textbf{Chair} \\
        \midrule
        Holodeck                     & 1.55 | 37.5\%       & 2.08 | 22.5\%        & 2.55 | 22.5\%       & --                 & --             \\
        PSDL                         & 1.52 | 42.5\%       & 0.72 | 55.0\%        & 0.33 | 90.0\%       & --                 & --             \\
        \midrule
        \methodname w/o user guidance       & 1.38 | 37.5\%       & 0.25 | 77.5\%        & 2.35 | 60.0\%       & \textbf{0.00 | 100\%}       & 0.10 | 95.0\%  \\
        \methodname w/o graph               & 2.60 | 15.0\%       & 1.74 | 36.8\%        & 6.94 | 38.9\%       & 2.63 | 68.4\%      & 0.35 | 95.0\%  \\
        \methodname w/o concepts            & 1.68 | 26.3\%       & 1.85 | 15.0\%        & 8.25 | 35.0\%       & 2.95 | 35.0\%      & 0.05 | 95.0\%  \\
        \midrule
        \methodname                         & \textbf{0.00 | 100\%} & \textbf{0.00 | 100\%} & \textbf{0.03 | 97.5\%} & \textbf{0.00 | 100\%} & \textbf{0.00 | 100\%} \\
        \bottomrule
    \end{tabular}}
\end{table}

%% file: appendix/app_D_experiment_details.tex
\section{Additional Experiment Details}
\label{sec:app_exp_details}

\subsection{Layout Program Generation Procedure}
\label{sec:app_layout_gen}

We use gpt-5.5 as the LLM for all layout-program generation experiments.
Each LLM call takes 17 seconds on average, and the correction loop runs for 1.2 rounds on average per generated program (capped at 2), at a mean cost of \$0.25 USD per program.
When prompting the LLM, we include in-context layout programs that the user previously marked as good for that domain.

\subsection{Prompt Sets}
\label{sec:app_prompt_sets}

For our standard single-scene generation use case, we use 20 prompts per domain.
We generate two samples per prompt for the comparison against Holodeck and PSDL in the dining, bedroom, and living room domains, and one sample per prompt for the ablation conditions.
To produce this set, we asked an LLM to generate prompts that exercise structural variations within each domain, given a few in-context examples of prompts that require compositional layouts.
Generic specifications, such as ``a bedroom,'' as used by prior work, are underspecified and not discriminative for our purpose.
Our goal is to evaluate how well a method aligns its output with the spatial patterns a user requests, and a generic prompt provides little concrete structure to evaluate.
For the living and bedroom domains we append the short instruction ``\textit{Add any furniture or decor needed to make the space feel complete and functional.}'' to each prompt.
Without this, we observed that LLMs tend to insert only the exact objects mentioned in the prompt, producing sparse rooms.
We do not append this instruction in the chair, bookshelves, or dining domains, where the prompts already specify the intended object set tightly.

We also construct a challenge set for the comparison between \methodname and PSDL in Section~\ref{sec:res_complexity}.
This set contains 30 additional prompts across the living room, bedroom, and dining domains that were written to be more difficult than the standard prompts.
The challenge prompts contain approximately 220 words and 18 explicitly requested constraints on average, compared with approximately 70 words and seven constraints in the standard set.
They produce layouts containing approximately 38 objects on average.
Section~\ref{sec:app_challenge_results} reports the results on this set.

\subsection{Comparison Methods}
\label{sec:app_comparisons}

For fairness, we use the same LLM (gpt-5.5, no reasoning) for both comparison methods and all of our own runs.
This is a more recent and capable model than the LLMs used in the original Holodeck and PSDL works, so our comparisons are, if anything, generous to the baselines.
Both baselines also explicitly try to avoid base validity errors.
Holodeck includes physical constraints in its optimization, while PSDL uses a correction loop to repair these errors.
Physical validity is therefore not a metric that only our solver optimizes for.

We run Holodeck out of the box on our prompt set.
A single layout takes around five minutes to generate.
Holodeck retrieves and arranges assets from the ObjaThor library and is tightly coupled to those assets for rendering.
We therefore cannot render Holodeck outputs through our final rendering pipeline (Section~\ref{sec:app_rendering}), and the qualitative comparisons display Holodeck's own ObjaThor renders directly.

PSDL produces a layout in three main steps: (i) from a text prompt, the LLM drafts a \emph{template} that declares the room dimensions and instantiates the assets to be placed, (ii) the LLM takes that template as input and writes a procedural program that lays out the assets, and (iii) the program is executed and solver violations are fed back through a correction loop.
Both PSDL and our method focus on the layout of entities, so we give both methods the same entities for each generation.
We bypass step (i) and, for each of our generated layouts, programmatically convert its solved scene representation into a PSDL template.
This template carries our domain's room dimensions and the same list of object instances as our output, including their categories and dimensions.
We then run PSDL's steps (ii) and (iii) on that template using the same text prompt as our run.
The full PSDL pipeline takes about 15 seconds per prompt, including the correction loop.
PSDL outputs are rendered through our final rendering pipeline so that the comparisons between program-based methods share the same visual style.

For the comparison of how well methods encode spatial patterns in Section~\ref{sec:res_encoding}, we select three concepts from each of the three comparison domains and write two layout prompts for each concept.
Because PSDL and Holodeck have no native way to align to a user's definition of a spatial pattern, we attempt to provide this by appending the user's concept description to the layout prompt.
In contrast, \methodname can use the corresponding library concept developed through our co-design interface.

\subsection{Quantitative Metrics}
\label{sec:app_metrics}

Our quantitative metrics count three general 3D scene validity properties: \textit{out-of-bounds} (an entity extends past the bounds of the room or container), \textit{intersection} (an entity overlaps another), and \textit{floating} (an entity is neither on the ground nor in stable contact with a supporting surface).
For 3D domains we score these directly using our base validity checks.
For 2D domains (dining, bookshelves), we score the same checks on the 3D scene produced by inserting per-category assets at the layout's predicted positions, so all methods are evaluated on a comparable 3D representation.

We use two sub-conditions for specific domains.
For the chair domain we relax both intersection and support: parts are allowed to interpenetrate, and a part is considered ``supported'' if it lies within $\varepsilon = 0.05$ of any other part rather than requiring a strict contact surface.
For Holodeck, the output contains many intersections between assets, some are visually valid (e.g., a cup resting on a shelf is recorded as intersecting that shelf), while others are genuine violations that would fail our full support check.
We do not attempt to distinguish these, thus we do not count any intersection between assets as a violation, so the violation counts we report for Holodeck are a lower bound on its true error rate.

To avoid potential confusion, we make clear the distinction between the terms \textit{valid} and \textit{solved} in our evaluation.
A valid layout, as reported in Figure~\ref{fig:validity_adherence} and Table~\ref{tab:domain_solve_comp}, passes every base validity check.
A solved layout, as reported in Table~\ref{tab:ablation}, also satisfies every concept verification in its generated concept-graph.
Solved is therefore the stricter criterion, which explains how the same method, for the same prompt set, can produce different valid and solved scores.

\textbf{Prompt adherence grading.}
For each prompt, we instruct an LLM to extract a checklist of the stated object-based constraints, such as ``the nightstand is beside the bed on its free side.''
These checklists then provide an automatic way to evaluate how well a layout matches the specifications of a layout prompt.
From the standard prompt set, we develop 81 such checks for living rooms, 77 for bedrooms, and 115 for dining arrangements.
We use Claude Opus 5 as the LLM for this judging, with low reasoning.
It receives the constraint checklist and a text description of the layout containing the room size and one line per object with its parameterization.
For each check, the judge returns a short justification and a true or false verdict.

\subsection{Extensions Setup}
\label{sec:app_extensions_setup}

\textbf{Generative layout programs (Section~\ref{sec:ext_metaprog}).}
We use 30 prompts, with 10 prompts from each of the living room, bedroom, and dining domains.
Each prompt pairs a scene description with two control parameters.
We sample two programs authored by the LLM per prompt and three example parameterizations per program.
We solve each combination of program and parameterization four times to test the robustness of these generations, and compute validity over every execution.
For adherence, we judge the execution with the fewest base validity violations for each parameterization and include the parameter values in the checklist shown to the judge.
We consider a program valid across its evaluated parameter range only if every execution is valid.

For this comparison, PSDL retains its own program authoring prompt and DSL.
We can no longer pass in a template parsed from \methodname's output, because the template would need to change depending on the parameterization of the layout program.
We instead provide the same per-domain object vocabulary used within \methodname.
We then ask PSDL to produce a function that takes the two controls as arguments and supplies three example parameterizations, matching the output format used by \methodname.
We pass the control values into PSDL as fixed inputs so its correction loop cannot change them, and then direct PSDL to select its own objects, dimensions, and room size while authoring these \metaprogs.

%% file: appendix/app_E_user_study.tex
\section{Non-Expert User Study Details}
\label{sec:app_user_study}

We recruited eight university students, each assigned to one of three domains: dining, living, or bedroom.
Sessions began with a $\sim$5-minute walkthrough of the interface: how the concept specification is entered, how scenarios and scenes can be inspected and edited, where text feedback is supplied, and how a concept is confirmed.
The study then proceeded in two parts.
In Part 1, each participant co-designed two concepts from short concept specifications written by an expert, progressing from an \textit{easier} concept to a \textit{harder} concept (as measured by expert effort).
In Part 2, four participants additionally proposed and co-designed a concept of their own.
For each concept, the concept specification was entered into the co-design interface to produce an initial implementation.
The user worked with the LLM to refine the concept's behavior to match the target intent, by providing guidance through the different interface views.
On average, developing a concept took $\sim$15 minutes, with $\sim$4.5 rounds of text guidance and $\sim$2.5 uses of \textit{auto-fix}.
After each concept, participants rated their agreement with the statement that the resulting concept matched the intended behavior; every rating was \emph{agree} or \emph{strongly agree}.
Sessions ended with a free-form interview.

Participants generally described the initial LLM implementation as a reasonable starting point that was easy to refine by inspecting scenarios, with one remarking that it ``tended to work pretty well'' and that it was straightforward to spot where it fell short.
In some cases the initial generation proposed a sensible interface that had not been asked for explicitly.
While a user was co-designing an \texttt{Align} concept, the LLM added a \texttt{spacing} parameter set to \texttt{equal}, which was outside the participant's original specification.
The user decided that this was a ``good idea'' and kept this functionality, noting that it injected a domain constraint they would not have written by hand.
When a verifier disagreed with their intent, participants sometimes wanted to give feedback, but found it hard to verbalize exactly what to change. 
In these cases they defaulted to labeling examples and invoking \textit{auto-fix}, as one participant did on the harder \texttt{CornerBedGroup} task.
In Part 2 of the study, participants reported that this process helped them settle on the design intent they wanted.
One participant reported not having a precise formulation of \texttt{Surround} in mind before starting.
For them, new requirements, such as the surrounding objects being evenly distributed, only surfaced once they saw the system's first set of scenarios.
Another participant described their \texttt{LookAt} session as a ``directed search'' through the concept space (Figure~\ref{fig:concept_evolution}, bottom, traces the development of this concept).

%% file: appendix/app_F_authoring_walkthrough.tex
\section{Layout Program Authoring Walkthrough}
\label{sec:app_walkthrough}

We walk through one run of the LLM pipeline for authoring layout programs from Section~\ref{sec:compose_concepts}.
We consider the following prompt:
\textit{``Generate a dining arrangement.
A tea-room round table for four where each place setting is reduced to a placemat with a cup and a spoon (no fork, no knife, no plate).
A single large candle sits in the exact center of the table.''}

Figure~\ref{lst:walkthrough_prompt} shows the prompt sent to the LLM.
For readability we elide three sections: the library API (interface and docstring of each of the dining concepts), a set of modeling guidelines on concept-graph best practices, and in-context layout program examples.
While most of the prompt is domain-general, some prompt material is taken from the base knowledge of the particular domain, including the domain definition, operations allowed on entities, and bounded space construction.

\begin{lstlisting}[style=apptext,caption={An example layout program authoring prompt for the dining domain.},label={lst:walkthrough_prompt}]
You are an expert procedural modeler, working in a structured visual domain composed of objects and spatial relationships.

## Domain Definition

You are designing dining arrangements by arranging objects in 2D space (top-down plan view), with optional vertical stacking via discrete levels.

In this coordinate space: the -X/+X axis is left/right, and the -Y/+Y axis is down/up.

Each dining region is modeled as a rectangle parameterized by a width and height. The scene bounds are:
  - X in [-width/2, +width/2]
  - Y in [-height/2, +height/2]

Objects are represented as oriented 2D bounding boxes (OBBs) with a center (x,y) position, width/height dimensions, and an orientation in the plane.

Objects can also be stacked using an integer 'level' attribute. Level 0 corresponds to objects supported by the ground (e.g., tables or furniture in the region). Higher levels represent objects supported by objects on the level below (e.g., plates on a table, utensils next to plates).

## Modeling Environment

Scenes in this domain are constructed using **ConceptGraphs**. A concept-graph consists of:

- **Nodes** — primitives or higher-level groupings that can be manipulated as a unit.
  - **Entity**: an atomic primitive defined by the domain.
  - **Group**: a higher-level grouping/layout operator over existing nodes.
  - A **child ConceptGraph** can itself be used as a Node inside a parent graph.

- **Relations** — constraints that relate nodes spatially or logically

When executed, a concept-graph synthesizes a concrete scene: a collection of objects arranged in space.

## API

{API DEF HERE}

## Modeling Guidelines:

{MODELING GUIDELINES HERE}

## Goal

Your job is to output **Python code only** that constructs a **scene-level** ConceptGraph using the above domain-specific API of Nodes, Relations, and Groups.

Do not render, simulate, or execute anything yourself.
Only use classes and parameters defined in the API. Do not invent new concepts or arguments.

### Construction pattern

```python
D = make_domain()
scene_dims = {'width': width, 'height': height}
CG = ConceptGraph(D, scene_dims)
```

* `CG.add_node(name, node)` adds an Entity or a child ConceptGraph.
* `CG.add_group(name, group)` introduces a Group.
* `CG.add_relation(name, Relation)` introduces a Relation.

---

## Hints

* Naming: names must be unique within a single ConceptGraph.
* Scoping: a Relation or Group may only reference names that exist in the **same** ConceptGraph where it is added.

## EXAMPLES

{IN CONTEXT EXAMPLES HERE}

## Output Requirements

* Output Python code only, which returns the scene-level ConceptGraph variable.
* Keep graphs readable: meaningful names, logical grouping, minimal redundancy.
* Do not mention runtime execution or internal implementation details.
* Follow the modeling guidelines. Carefully adhere to both their rules and best-practice recommendations.

## Task

Use this API to define a scene with the following description: "Generate a dining arrangement. A tea-room round table for four where each place setting is reduced to a placemat with a cup and a spoon (no fork, no knife, no plate). A single large candle sits in the exact center of the table."

\end{lstlisting}

Figure~\ref{lst:walkthrough_v1} is the LLM's response.
It builds a child concept-graph for a place setting containing a placemat, a cup supported on the placemat, and a spoon to the right of and near the cup.
In the top-level concept-graph, the \texttt{CircularTableLayout} \nodename copies the setting and a chair four times around a round table.
A \texttt{Supporting} \edgename centers a candle on the table.

\begin{lstlisting}[style=apppython,caption={The LLM's first layout program.},label={lst:walkthrough_v1}]
def make_scene():
    D = make_domain()
    import random

    def pt(a, b=0.03):
        return random.uniform(a - b, a + b)

    scene_dims = {"width": pt(5.5, 0.4), "height": pt(5.5, 0.4)}
    scene = ConceptGraph(D, scene_dims, bound_mode="hard")

    # -----------------------------
    # Minimal tea place setting:
    # placemat with a cup and spoon only.
    # -----------------------------
    setting_w = pt(0.55, 0.04)
    setting_h = pt(0.45, 0.03)
    setting = ConceptGraph(
        D,
        {"width": setting_w, "height": setting_h},
        bound_mode="hard",
    )

    setting.add_node(
        "placemat",
        Placemat(width=setting_w, height=pt(0.30, 0.02)),
    )
    setting.add_node(
        "cup",
        Cup(width=pt(0.12, 0.01), height=pt(0.12, 0.01)),
    )
    setting.add_node(
        "spoon",
        Spoon(width=pt(0.045, 0.005), height=pt(0.24, 0.02)),
    )

    setting.add_relation(
        "cup_on_placemat",
        Supporting(node_name_list=["cup", "placemat"], centered=True),
    )
    setting.add_relation(
        "spoon_right_of_cup",
        ToRight(node_name_list=["spoon", "cup"]),
    )
    setting.add_relation(
        "spoon_near_cup",
        Near(node_name_list=["spoon", "cup"], distance=pt(0.08, 0.02)),
    )

    # -----------------------------
    # Round tea table for four.
    # CircularTableLayout duplicates the setting and chair prototypes evenly
    # around the table.
    # -----------------------------
    scene.add_node(
        "tea_table",
        RoundTable(width=pt(2.2, 0.10), height=pt(2.2, 0.10)),
    )
    scene.add_node("tea_setting_proto", setting)
    scene.add_node(
        "chair_proto",
        Chair(width=pt(0.55, 0.04), height=pt(0.55, 0.04)),
    )

    scene.add_group(
        "tea_table_for_four",
        CircularTableLayout(
            node_name_list=["tea_table", "tea_setting_proto", "chair_proto"],
            num_settings=4,
        ),
    )

    scene.add_relation(
        "tea_table_centered",
        Centered(node_name_list=["tea_table"], max_center_dist=pt(0.12, 0.02)),
    )

    # -----------------------------
    # Single large candle at the exact center of the table.
    # -----------------------------
    scene.add_node(
        "center_candle",
        Candle(width=pt(0.28, 0.02), height=pt(0.28, 0.02)),
    )
    scene.add_relation(
        "candle_centered_on_table",
        Supporting(node_name_list=["center_candle", "tea_table"], centered=True),
    )

    return scene
\end{lstlisting}

\begin{figure}[t!]
\centering
\includegraphics[width=0.7\linewidth]{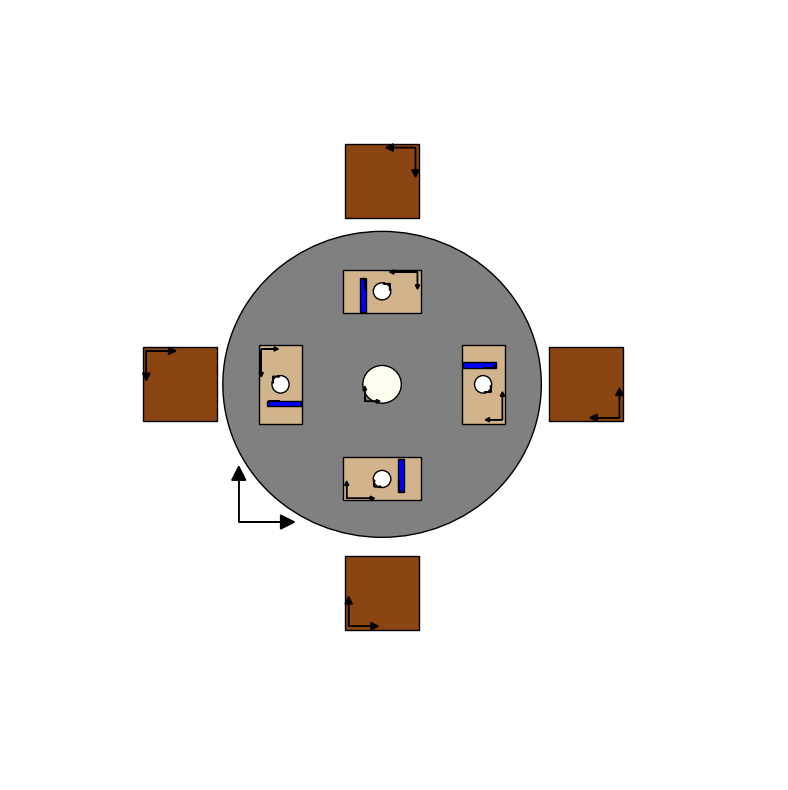}
\caption{Best layout found for the first program.
Each copied spoon overlaps its placemat.}
\label{fig:walkthrough_v1}
\end{figure}

Within the time budget, our solver could not find a solution with zero error to the concept-graphs produced by this layout program.
Figure~\ref{fig:walkthrough_v1} shows its best layout (in terms of graph error).
The arrangement is reasonable, but in every copied setting the spoon and the placemat both rest on the table at the same stacking level, and placing the spoon beside the cup puts it on top of the placemat.
The base validity checks flag this as an overlap.
Figure~\ref{lst:walkthrough_feedback} shows the text returned to the LLM: the base validity error (reported once per copied setting) followed by instructions on how to treat the failure list.
The LLM receives this text in the same conversation as the original prompt and its program.

\begin{lstlisting}[style=apptext,float=tb,caption={The solver feedback sent back to the LLM.},label={lst:walkthrough_feedback}]
The solver was unable to find any satisfying layouts for the ConceptGraphs generated by your program.
Here are some of the errors from the best run the solver found:

~~ ConceptGraph -- scene ~~ 
  Base verification error : Objects 2 ('placemat') and 3 ('spoon') overlap at level 1.
  [... the same error is reported for the other three copied settings ...]

## Iteration guidance

Your aim is to converge to a concept-graph that is both *satisfiable* and that fully captures the user's prompt. Treat the failure list as a signal to repair, not a delete-list.

* A failing concept may simply have parameter values that are slightly off — updating them is often enough to make it pass.
* When a concept fails, look at the full set of concepts that touch any of its nodes. The flagged concept may be correct, and an unflagged neighbor may be the one creating the conflict — consider changing or removing the neighbor instead.
* Only delete a concept when its conflict with the rest of the graph is genuinely unavoidable. Deleting on first failure tends to discard real intent and leaves the scene under-constrained.
* If the current graph does not yet express some part of the user's prompt, add concepts to fill the gap. Iterations are not limited to editing what is already there — the graph should carry enough constraints to match the prompt's intent.
\end{lstlisting}

\begin{figure}[t!]
\centering
\includegraphics[width=0.48\linewidth]{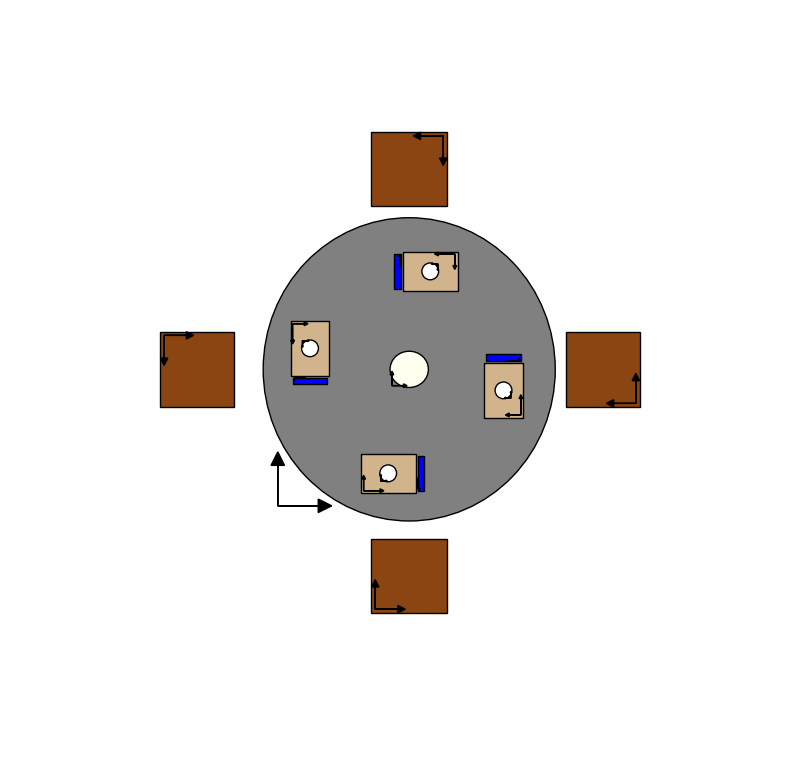}\hfill
\includegraphics[width=0.48\linewidth]{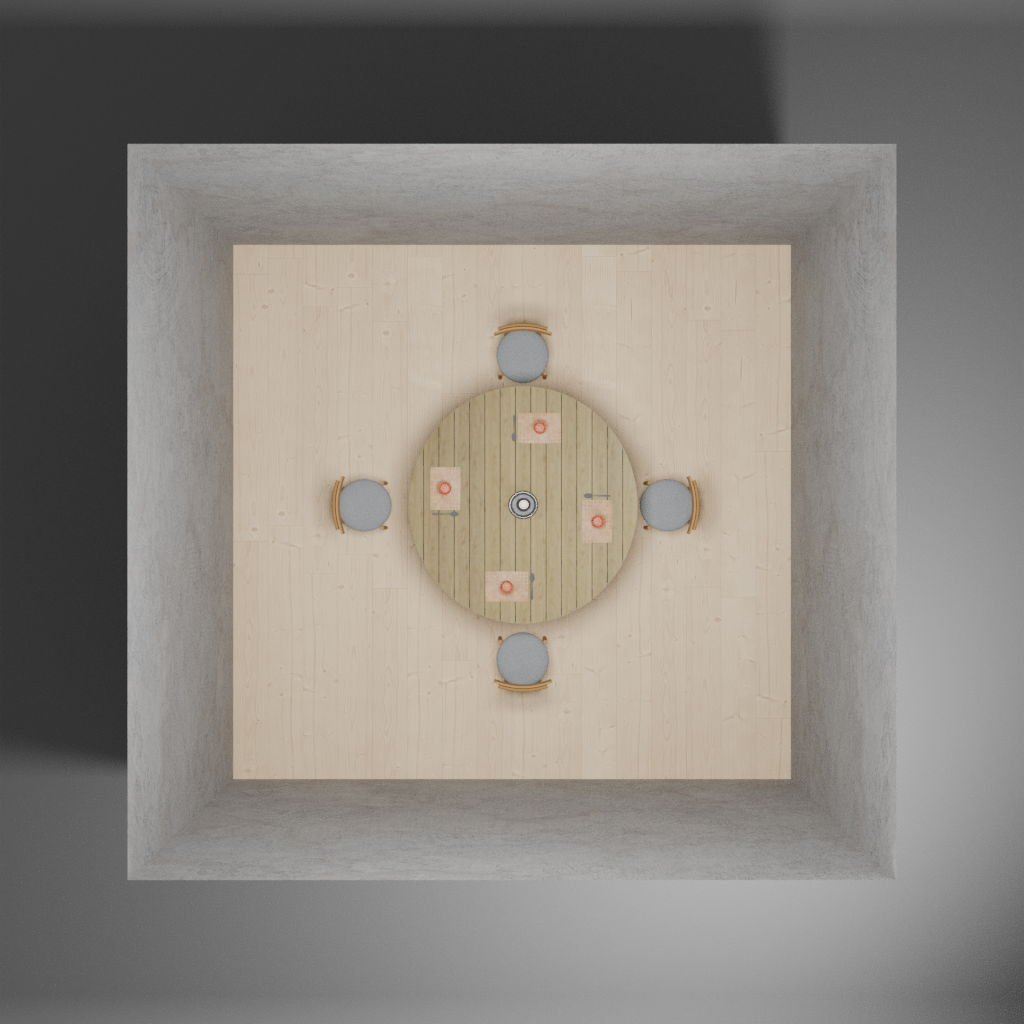}
\caption{Layout produced by the repaired program: structured render (left) and final render (right).}
\label{fig:walkthrough_v2}
\end{figure}

Figure~\ref{lst:walkthrough_diff} shows the lines the LLM changed.
It retargets the spoon's \texttt{ToRight} and \texttt{Near} \edgenames from the cup to the placemat, adds an \texttt{Aligned} \edgename so the spoon sits level with the placemat, widens the setting, and narrows the placemat.
The table, chairs, candle, and \texttt{CircularTableLayout} \nodename are unchanged.
The solver can then find valid solutions to concept-graphs produced from the repaired program.
Figure~\ref{fig:walkthrough_v2} shows the resulting structured layout and its final render.

\begin{lstlisting}[style=appdiff,caption={Changed lines between the first and the repaired program.},label={lst:walkthrough_diff}]
--- program v1
+++ program v2
@@ -11,8 +11,10 @@
     # -----------------------------
     # Minimal tea place setting:
     # placemat with a cup and spoon only.
+    # The cup is supported on the placemat; the spoon is kept adjacent to the
+    # placemat rather than on top of it to avoid same-level overlap.
     # -----------------------------
-    setting_w = pt(0.55, 0.04)
+    setting_w = pt(0.75, 0.04)
     setting_h = pt(0.45, 0.03)
     setting = ConceptGraph(
         D,
@@ -22,7 +24,7 @@
 
     setting.add_node(
         "placemat",
-        Placemat(width=setting_w, height=pt(0.30, 0.02)),
+        Placemat(width=pt(0.42, 0.03), height=pt(0.30, 0.02)),
     )
     setting.add_node(
         "cup",
@@ -38,12 +40,16 @@
         Supporting(node_name_list=["cup", "placemat"], centered=True),
     )
     setting.add_relation(
-        "spoon_right_of_cup",
-        ToRight(node_name_list=["spoon", "cup"]),
+        "spoon_right_of_placemat",
+        ToRight(node_name_list=["spoon", "placemat"]),
     )
     setting.add_relation(
-        "spoon_near_cup",
-        Near(node_name_list=["spoon", "cup"], distance=pt(0.08, 0.02)),
+        "spoon_near_placemat",
+        Near(node_name_list=["spoon", "placemat"], distance=pt(0.08, 0.02)),
+    )
+    setting.add_relation(
+        "spoon_aligned_with_placemat",
+        Aligned(node_name_list=["spoon", "placemat"], axis="horizontal"),
     )
 
     # -----------------------------
\end{lstlisting}

%% file: appendix/app_G_library.tex
\section{Concept Library Details}
\label{sec:app_library}

The five expert co-designed libraries contain over 90 concepts with over 18,000 lines of code and comments.
Here, for demonstration, we show the full logic of the \texttt{Surround} concept (discussed in Section~\ref{sec:CD_walkthrough}).

\begin{lstlisting}[style=apppython,caption={The implementation of the \texttt{Surround} library concept.},label={lst:surround}]
class Surround(Relation):
    def __init__(self, node_name_list: list[str], distance: float, spacing: str):
        """
        Constrains a central node to be surrounded by multiple outer nodes.

        Parameters
        ----------
        node_name_list : list[str]
            A list of at least four node names:
            [center_node, outer_node_1, outer_node_2, ...]. The first entry
            is treated as the center node; all remaining entries are treated
            as outer nodes that should surround the center.
        distance : float
            Target radial distance from the center node to each outer node,
            interpreted in scene units. During synthesis, outer nodes should
            be placed approximately this far from the center; verification
            should check that this relationship is approximately satisfied.
        spacing : str
            Controls angular spacing between outer nodes. When set to
            'even', outer nodes are expected to be arranged with uniform
            angular separation around the center. When set to 'uneven',
            outer nodes may have non-uniform angular spacing and may vary
            more in their exact distance from the center.
        """
        if not isinstance(node_name_list, list) or len(node_name_list) < 4:
            raise ValueError(
                "Surround requires node_name_list to contain at least four "
                "elements: [center_node, outer_node_1, outer_node_2, ...]."
            )

        if spacing not in ("even", "uneven"):
            raise ValueError("Surround 'spacing' must be either 'even' or 'uneven'.")

        # Store parameters for later synthesis and verification
        self.node_name_list = node_name_list
        self.distance = float(distance)
        self.spacing = spacing

        # Derived convenience references (positional only; names are opaque)
        self.center_node_name = node_name_list[0]
        self.outer_node_names = node_name_list[1:]

        # Call base Relation initializer with all parameters unchanged
        super().__init__(node_name_list=node_name_list,
                         distance=distance,
                         spacing=spacing)

    # -- Synthesis --
    def Synthesis(self, obj_list):
        import math
        import random

        # Guard: need at least one center and one outer node
        if len(obj_list) < 2:
            return obj_list

        center = obj_list[0]
        outers = obj_list[1:]
        n = len(outers)

        if n == 0:
            return obj_list

        cx, cy = center.get_position()
        base_radius = max(float(self.distance), 0.0)

        # Default to even if spacing is missing or invalid
        spacing_mode = getattr(self, "spacing", "even")
        even_spacing = spacing_mode == "even" or n == 1

        full_circle = 2.0 * math.pi

        def orient_outer_toward_center(outer, ox, oy):
            # Compute vector from outer to center
            vx = cx - ox
            vy = cy - oy
            dist_sq = vx * vx + vy * vy
            if dist_sq == 0.0:
                return  # Degenerate; skip orientation

            desired_angle = math.atan2(vy, vx)

            dirs = outer.get_dirs()
            fy, fx = dirs["y_dir"][1], dirs["y_dir"][0]  # current +y_dir
            curr_angle = math.atan2(fy, fx)

            delta_rad = desired_angle - curr_angle
            delta_deg = math.degrees(delta_rad)
            outer.rotate_about_self(delta_deg)

        if even_spacing:
            # Uniform angular spacing around full circle
            angle_step = full_circle / n
            start_angle = 0.0

            for i, outer in enumerate(outers):
                angle = start_angle + i * angle_step
                x = cx + base_radius * math.cos(angle)
                y = cy + base_radius * math.sin(angle)
                outer.set_pos(x, y)
                orient_outer_toward_center(outer, x, y)
        else:
            # Uneven: jitter both angle and radius slightly
            angle_step = full_circle / n
            base_angles = [i * angle_step for i in range(n)]

            # Jitter each angle within ±40% of the nominal step (keeps order)
            max_angle_jitter = 0.4 * angle_step
            jittered_angles = [
                a + random.uniform(-max_angle_jitter, max_angle_jitter)
                for a in base_angles
            ]

            for angle, outer in zip(jittered_angles, outers):
                # Radial jitter: ±20% around the base distance
                radius = base_radius * random.uniform(0.8, 1.2)
                x = cx + radius * math.cos(angle)
                y = cy + radius * math.sin(angle)
                outer.set_pos(x, y)
                orient_outer_toward_center(outer, x, y)

        return obj_list

    # -- Verification --
    def Verification(self, obj_list):
        import math
        from typing import List

        messages: List[str] = []

        # Basic arity checks
        expected_n = len(self.node_name_list)
        if len(obj_list) != expected_n:
            messages.append(
                f"Surround Relation expects {expected_n} object groups (for nodes "
                f"{self.node_name_list}), got {len(obj_list)}."
            )
            return False, messages

        if expected_n < 4:
            messages.append(
                "Surround requires at least four nodes "
                "[center_node, outer_node_1, outer_node_2, ...]."
            )
            return False, messages

        center_group = obj_list[0]
        outer_groups = obj_list[1:]
        outer_names = self.outer_node_names

        if len(outer_groups) != len(outer_names):
            messages.append(
                "Mismatch between number of outer groups and outer node names in Surround."
            )
            return False, messages

        # Get center position
        cx, cy = center_group.get_position()

        # Collect positions, radii, and angles of all outer nodes relative to the center
        radii: List[float] = []
        angles: List[float] = []
        outer_positions: List[tuple[float, float]] = []

        for og in outer_groups:
            x, y = og.get_position()
            dx = x - cx
            dy = y - cy
            r = math.hypot(dx, dy)
            theta = math.atan2(dy, dx)  # in [-pi, pi]
            outer_positions.append((x, y))
            radii.append(r)
            angles.append(theta)

        # --- Radial distance checks ---
        target_r = float(self.distance)
        eps = 1e-6

        # Only enforce a radial target if it is positive
        if target_r > 0.0:
            if self.spacing == "even":
                rel_tol = 0.25  # 25% relative tolerance
                abs_tol = 0.05  # absolute fallback tolerance
            else:  # "uneven"
                rel_tol = 0.5   # 50% relative tolerance
                abs_tol = 0.1

            tol_r = max(abs_tol, rel_tol * target_r)

            for name, r in zip(outer_names, radii):
                if abs(r - target_r) > tol_r + eps:
                    messages.append(
                        f"Surround radial distance violated for outer node '{name}' "
                        f"around center '{self.center_node_name}': "
                        f"radius {r:.4f} differs from target {target_r:.4f} by "
                        f"more than allowed {tol_r:.4f}."
                    )

        # --- Angular coverage and spacing checks ---
        m = len(outer_groups)
        if m >= 2:
            # Sort by angle to examine angular gaps
            sorted_indices = sorted(range(m), key=lambda i: angles[i])
            sorted_angles = [angles[i] for i in sorted_indices]

            gaps: List[float] = []
            for i in range(m - 1):
                gaps.append(sorted_angles[i + 1] - sorted_angles[i])
            # Wrap-around gap
            wrap_gap = (sorted_angles[0] + 2.0 * math.pi) - sorted_angles[-1]
            gaps.append(wrap_gap)

            max_gap = max(gaps)

            # Coverage check: ensure no single gap exceeds 180 degrees
            if max_gap > math.pi + eps:
                messages.append(
                    f"Surround coverage violated around center '{self.center_node_name}': "
                    f"largest angular gap is {math.degrees(max_gap):.1f}° (> 180°), "
                    "so the center is not well surrounded by outer nodes."
                )

            # Even spacing check (only if requested)
            if self.spacing == "even":
                target_gap = 2.0 * math.pi / m
                tol_gap = 0.35 * target_gap  # allow ±35% variation

                for gi, gap in enumerate(gaps):
                    if abs(gap - target_gap) > tol_gap + eps:
                        i0 = sorted_indices[gi]
                        i1 = sorted_indices[(gi + 1) % m]
                        name0 = outer_names[i0]
                        name1 = outer_names[i1]
                        messages.append(
                            f"Surround even spacing violated between outer nodes "
                            f"'{name0}' and '{name1}' around center "
                            f"'{self.center_node_name}': angular gap "
                            f"{math.degrees(gap):.1f}° deviates from expected "
                            f"{math.degrees(target_gap):.1f}° by more than "
                            f"{math.degrees(tol_gap):.1f}°."
                        )

        # --- Orientation / facing checks ---
        # Outer nodes should "face" the center. We interpret the local +y_dir as the
        # facing direction and require it to point approximately toward the center.
        facing_cos_threshold = math.cos(math.radians(60.0))  # ~0.5

        for og, (ox, oy), r, name in zip(outer_groups, outer_positions, radii, outer_names):
            # If an outer node is essentially at the center, skip facing check
            if r <= eps:
                continue

            dirs = og.get_dirs()
            y_dir = dirs.get("y_dir", None)
            if y_dir is None or len(y_dir) != 2:
                # If we cannot determine orientation, report a violation
                messages.append(
                    f"Surround facing violated for outer node '{name}' around center "
                    f"'{self.center_node_name}': missing or invalid 'y_dir' orientation."
                )
                continue

            fyx, fyy = y_dir[0], y_dir[1]
            norm_f = math.hypot(fyx, fyy)
            if norm_f <= eps:
                messages.append(
                    f"Surround facing violated for outer node '{name}' around center "
                    f"'{self.center_node_name}': 'y_dir' orientation has near-zero length."
                )
                continue

            # Vector from outer node to center
            vx = cx - ox
            vy = cy - oy
            norm_v = math.hypot(vx, vy)
            if norm_v <= eps:
                # Already handled by r <= eps above, but keep the guard
                continue

            # Cosine of angle between facing direction and vector to center
            dot = (vx * fyx + vy * fyy) / (norm_v * norm_f)

            if dot < facing_cos_threshold:
                angle_deg = math.degrees(math.acos(max(-1.0, min(1.0, dot))))
                messages.append(
                    f"Surround facing violated for outer node '{name}' around center "
                    f"'{self.center_node_name}': facing direction is at "
                    f"{angle_deg:.1f}° from the direction to the center, which is "
                    f"greater than allowed 60.0°."
                )

        is_valid = len(messages) == 0
        return is_valid, messages
\end{lstlisting}